\documentclass[structabstract]{aa}  
\usepackage{graphicx}
\usepackage{txfonts}
\begin{document}
   \title{The role of major mergers in shaping galaxies at $2\lesssim z<4$ in 
   the VUDS and VVDS surveys}

   \author{L. A. M.~Tasca\inst{1}
	  \fnmsep\thanks{Based on data obtained with the European 
	  Southern Observatory Very Large Telescope, Paranal, Chile, under Large
	  Programs 070.A--9007, 177.A--0837 and 185.A--0791. 
	  Based on observations obtained with MegaPrime/MegaCam, a joint project
	  of CFHT and CEA/DAPNIA, at the Canada--France--Hawaii Telescope (CFHT)
	  which is operated by the National Research Council (NRC) of Canada, 
	  the Institut National des Sciences de l'Univers of the Centre National de la Recherche
          Scientifique (CNRS) of France, and the University of Hawaii. This work
	  is based in part on data products produced at TERAPIX and the
          Canadian Astronomy Data Centre as part of the Canada--France--Hawaii
          Telescope Legacy Survey, a collaborative project of NRC and CNRS.
          This work is based on observations taken by the CANDELS Multi--Cycle 
	  Treasury Program with the NASA/ESA HST, which is operated by the 
	  Association of Universities for Research in Astronomy, Inc., under 
	  NASA contract NAS5--26555.
	  }
          \and O.~Le F\`evre \inst{1}
	  \and C.~L\'opez--Sanjuan \inst{2}
	  \and P.-W.~Wang \inst{1}
	  \and P.~Cassata \inst{1}
  	  \and B.~Garilli \inst{3}
          \and O.~Ilbert \inst{1}
 	  \and V.~Le~Brun \inst{1}
	  \and B.~C.~Lemaux \inst{1}
  	  \and D.~Maccagni \inst{3}
 	  \and L.~Tresse  \inst{1}
 	  \and S.~Bardelli \inst{4}
     	  \and T.~Contini \inst{5,6}
 	  \and O.~Cucciati \inst{4}
	  \and A.~Fontana \inst{7}
	  \and M. Giavalisco \inst{8}
	  \and J.-P.~Kneib \inst{1,9}	  
 	  \and M.~Salvato \inst{10}
	  \and Y.~Taniguchi \inst{11}
	  \and D.~Vergani \inst{12}
 	  \and G.~Zamorani \inst{4}
  	  \and E.~Zucca \inst{4}	  
          }

   \institute{Aix Marseille Universit\'e, CNRS, LAM (Laboratoire d'Astrophysique 
   de Marseille) UMR 7326, 13388, Marseille, France \\
              \email{lidia.tasca@oamp.fr}
   \and
   Centro de Estudios de F\'isica del Cosmos de Arag\'on, Plaza San Juan 1, planta 2, 44001 Teruel, Spain
   \and
   INAF--IASF Milano, Milano, Italy
   \and
   INAF Osservatorio Astronomico di Bologna, via Ranzani 1, I-40127, Bologna, Italy
   \and
   Institut de Recherche en Astrophysique et Planetologie, CNRS, 14, avenue Edouard Belin, F--31400 Toulouse, France
   \and
   IRAP, Universite de Toulouse, UPS--OMP, Toulouse, France
   \and
   INAF - Osservatorio Astronomico di Roma Via Frascati 33 00040 Monteporzio (RM)
   \and
   University of Massachussetts
   \and
   LASTRO, Ecole polytechnique f\'ed\'erale de Lausanne, Suisse
   \and
   Max--Planck--Institut fur extraterrestrische Physik, Giessenbachstrasse, D--85748 Garching bei Munchen, Germany
   \and
   Research Center for Space and Cosmic Evolution, Ehime University, Bunkyo--cho, Matsuyama 790--8577, Japan
   \and 
   INAF--IASF Bologna, Via P. Gobetti 101, I--40129 Bologna, Italy
             \\
             }

   \date{Received March, 2013; accepted ..., 2013}

 
  \abstract
   {The mass assembly of galaxies can proceed through different physical 
   processes.
   Here we report on the spectroscopic identification of close physical pairs 
   of galaxies at redshifts $2 \lesssim z<4$ and discuss the impact of major 
   mergers in shaping galaxies at these early cosmological times.}
   {We aim to identify and characterize close physical pairs of galaxies 
   destined to merge and use their properties to infer the contribution of 
   merging processes to the early mass assembly of galaxies.}
   {We search for galaxy pairs with a transverse separation $r_p \leq 25$ 
   h$^{-1}$kpc and a velocity difference $\Delta v \leq 500$ km s$^{-1}$ using 
   the VIMOS VLT Deep Survey (VVDS) and early data from the VIMOS Ultra Deep 
   Survey (VUDS) that comprise a sample of $1111$ galaxies with spectroscopic 
   redshifts measurements at redshifts $1.8 \leq z \leq 4$ in the COSMOS, ECDFS, 
   and VVDS--02h fields.
   We analyse their spectra and associated visible and near--infrared photometry 
   to assess the main properties of merging galaxies that have an average stellar 
   mass $M_{\star}=2.3 \times 10^{10}$ M$_{\sun}$ at these redshifts.}
   {Using the 12 physical pairs found in our sample we obtain a first estimate 
   of the merger fraction at these rdshifts, $f_{merg} \sim(15-20)$\%. 
   The pair separations indicate that these pairs will merge within 1 Gyr, on 
   average, with each producing a more massive galaxy by the time  the cosmic 
   star formation peaks at $z\sim1-2$. 
   From the average mass ratio between galaxies in the pairs, the stellar mass 
   of the resulting galaxy after merging will be $\sim60\%$ larger
   than the most massive galaxy in the pair before merging. 
   We conclude that major merging of galaxy pairs 
   is on--going at $2\lesssim z<4$ and significantly 
   contributing to the major assembly phase of galaxies at this early epoch. }
   {}


   \maketitle
%

\section{Introduction}

The contribution of different physical processes to galaxy mass assembly along 
cosmic time is still unknown, and a clear picture describing 
how galaxies assemble, supported by observational evidence, has yet to emerge.
Looking back at the average history of a galaxy observed today, we are yet 
unable to identify how its stellar mass has been acquired and which physical 
processes are possible contributors.

The mass build--up of galaxies is expected to proceed through a relatively small number 
of processes (for a summary, see e.g. Springel et al. 2005a; Silk \& Mamon, 2012). 
New stars can form from the gas reservoir of a galaxy, either acquired at birth 
or replenished from a more continuous accretion process along the galaxy 
lifespan since formation. 
Major and minor merging between galaxies is identified in numerous spectacular 
examples in the local universe 
(e.g. in the RNGC catalogue, Sulentic \& Tifft, 1973; 
Barnes and Hernquist, 1992).
Merging is efficient at assembling mass, as it produces a significant increase 
in mass of up to a factor of two for equal mass mergers, for each merging event.  
Other processes are expected to modulate the total mass gains. AGN and SNe 
feedback have been proposed as mechanisms capable of quenching star 
formation, 
as well as supporting winds capable at driving some mass fraction into the 
inter--galactic medium (IGM), hence  reducing the increase in stellar mass 
from in--situ star formation (Silk, 1997; Murray et al. 2005; 
Cattaneo et al. 2009). 
The environment of galaxies is also
expected to impact mass growth, with interactions between galaxies and the dense
intra--cluster medium, like harassment or stripping, 
rapidly removing a significant part of the gas 
content of a galaxy (e.g. Moore et al. 1996).  
These processes are expected to ultimately combine along cosmic time to produce
the mass distribution observed in the well--defined Hubble sequence of galaxy 
types in the nearby  universe. 

In recent years, cold gas accretion fuelling star formation has received a
focused attention, following numerical simulations (Kere\v{s} et al. 2005). 
In this picture, cold gas flows along the filaments of the cosmic web 
into the main body of a galaxy to support vigorous star formation. 
This mechanism has  been proposed as the main mode of galaxy assembly 
(Dekel et al. 2009), and is often cited in the recent literature as the 
preferred scenario for galaxy assembly 
(e.g. Kere\v{s} et al. 2009; Dijstra \& Loeb, 2009; Bouch\'e et al. 2010;
di Matteo et al. 2012).
However, as of today, only limited and indirect observational evidence exist in 
support of this picture (Cresci et al. 2010; Kacprzak et al. 2012), while 
detailed observational investigations have failed to identify direct 
supporting evidence (Steidel et al. 2010), which demonstrates the difficulty in 
directly identifying the accretion process at work.

The merging of galaxies is another key process that contributes to galaxy 
assembly.
The hierarchical growth of dark matter halos is a key prediction of the 
$\Lambda CDM$ model for galaxy formation 
(Davis et al. 1985; Springel et al. 2005b; Hopkins et al. 2006). 
In this picture, the merging of dark matter (DM) halos would not only lead to
an increase of the DM halo masses, but also naturally lead to the merging of 
the galaxies associated with each of the halos 
(Kauffmann, White \& Guiderdoni 1993). 
While galaxy assembly seems to produce the more massive galaxies early in a 
seemingly anti--hierarchical downsizing pattern (De Lucia et al. 2006),
it is nonetheless expected that merging of galaxies would continue to occur as 
DM halos continue to merge along cosmic time.  
As the dynamical time--scale for halos to merge is of the order of 0.5--1 Gyr 
(Kitzbichler \& White 2008; Lotz et al. 2010), it is expected that a massive
halo today, as identified in DM halo merger trees from numerical N--body 
simulations, will have experienced several mergers since its formation.
Documenting merging activity at different epochs can therefore shed light 
on the contribution of this process in assembling mass in galaxies. 

Evidence for merging is  direct and well documented. 
Mergers have been identified since early days of photographic galaxy atlases 
and classified alongside the Hubble sequence of morphological types (e.g. the RC3 
catalogue, de Vaucouleurs et al. 1991). 
The merging process of two disc galaxies has been proposed as one of the 
mechanism for producing early--type bulge--dominated galaxies, supported from 
simulations (e.g. Mihos \& Hernquist 1996; Bournaud et al. 2005). 
Although this picture may be too simplistic (Bournaud et al. 2011), it is, 
however, clear that mergers do occur in the low redshift universe and that 
major mergers may lead to large modifications of the physical properties of the 
galaxies involved. 
Mergers are identified either {\it aposteriori} morphologically
from wisps, tails, or irregular shapes produced by on--going or 
post--merger dynamics, or {\it apriori} from the identification of 
pairs of physically bound galaxies destined to coalesce. 
Going to high redshifts, using pairs in  early merging stages, rather
than post--merger remnants, remains the most 
robust way to derive a merger fraction (e.g. Le F\`evre et al. 2000). 
This is because major merging pairs are easier to identify at these redshifts
than the post--merger morphological signatures, which are of low surface 
brightness and may escape detection. 
The pair fraction can be transformed into a merger rate per volume or per galaxy 
(Patton et al. 2000; Kitzbichler \& White 2008; Lin et al. 2008; 
de Ravel et al. 2009; Lopez--Sanjuan et al. 2011; Lopez--Sanjuan et al. 2013), 
using the 
dynamical time--scale for a pair of galaxies with a given mass ratio and 
projected physical separation (Kitzbichler and White 2008; Lotz et al. 2010). 
The integrated merging rate over the lifespan of a galaxy since formation would then give the 
total amount of stellar mass assembled from the merger process.
The secure identification of pairs and pair fractions at different redshifts is therefore 
an important observational measurement to perform. Spectroscopic redshift 
measurements of both members of the pair are required to eliminate the risk of 
background or foreground contamination along the line of sight and confirm
that the pair is physically bound. 

At $z \sim 1$ pairs have been observed from deep galaxy spectroscopic surveys 
(e.g. Lin et al. 2008; de Ravel et al. 2009).
From the VIMOS VLT Deep Survey (VVDS), de Ravel et al. (2009) found that the 
merger fraction is higher by a factor $\sim3$ at $z = 1$ than in the local 
Universe and further showed that the merger rate and its evolution depends 
significantly on the stellar mass (luminosity) of the galaxy population. 
At redshifts beyond $z\sim 1$ only a few direct identifications of pairs and
measurements of the merger fraction and merger rate exist. 
Lopez--Sanjuan et al. (2013) have reported a high pair fraction of 
$\sim20-22$\% at $1 < z < 1.8$ from 3D spectroscopy measurements in MASSIV  
(Mass Assembly Survey with Sinfoni in VVDS, Contini et al. 2012). 
At higher redshifts Conselice et al. (2003) have used the CAS (concentration, 
asymmetry clumpiness)  methodology, which relyies on image shapes and the 
expected signature of on--going or past mergers, to perform a 
measurement of the merger fraction up to $z\sim3$. 
Cooke et al. (2010) have provided spectroscopic identification of 5 pairs of 
galaxies in their LBG sample at $z \sim 3$, claiming that merging is triggering a significant
part of the Ly$\alpha$ emission. 
The number of confirmed pairs is therefore 
small beyond $z\sim2$, and larger samples have to be assembled to 
enable measurements of the pair fraction and merger rate which is accurate to a 
few percent.

Here we present a sample of galaxy pairs identified at $z>1.8$ in the VUDS 
(VIMOS Ultra Deep Survey) and VVDS. 
The VUDS is an on--going survey (Le F\`evre et al. 2013b, in prep.)
with ultra--deep spectroscopy obtained with VIMOS on the VLT targeting 
galaxies with $z>2$ in three well studied fields: 
the COSMOS, ECDFS and VVDS--02h (XMM--LSS/CFHTLS--D1).
The VVDS has been extensively discussed elsewhere (Le F\`evre et al. 2005). 
We are using the ``Final Data Release'' of this survey as described in 
Le F\`evre et al. (2013a, in prep.).
These spectroscopic redshift survey samples are searched to identify a sample 
of real physical pairs with redshifts $1.8 < z < 4$, based on the observed 
projected separation $r_p$ and velocity difference $\Delta$v. 
We discuss the derived pair properties using all available spectroscopy
as well as visible and near--IR imaging and photometric data.
The VUDS and VVDS spectroscopic redshift surveys are described in Section 
\ref{obs}, and the methodology to identify pairs and confirm that they are
at close physical separation rather than a random projection along the
line of sight is presented in Section \ref{pair_select}.
We then examine the pair properties in Section \ref{pairs_prop}. We discuss
the first detection of a significant number of major merging pairs at 
$1.8 < z < 4$ and conclude in Section \ref{discuss}.

Throughout this work, we adopt the ``convergence cosmology'' 
$H_0=100h~km~s^{-1}~Mpc^{-1}$, {\bf{$h=0.7$}},
$\Omega_{0,\Lambda}=0.73$ and $\Omega_{0,m}=0.27$. 
All magnitudes are given in the AB system.


\section{Spectroscopic observations}
\label{obs}

To look for pairs, we have explored the VUDS, VVDS--Deep, and VVDS--UltraDeep  
surveys, providing a sample of galaxies with spectroscopic redshifts measured 
with VIMOS on the ESO--VLT (Le F\`evre et al. 2003). 
The VIMOS spectra have been obtained with 4.5h of integration for the VVDS--Deep
survey,
covering $5500 \leq \lambda \leq 9350$\AA, and 16h and 14h integrations in 
each of the LRBLUE and LRRED grism settings for the VVDS-UltraDeep and VUDS
surveys respectively, covering a combined wavelength range 
$3600 \leq \lambda \leq 9350$\AA. The redshift accuracy with this
setup is $\simeq0.0005 \times (1+z)$ (Le F\`evre et al. 2013a).

The VVDS--deep (Le F\`evre et al. 2005; Le F\`evre et al. 2013a) and
VVDS--UltraDeep (Le F\`evre et al. 2013a) surveys are based on $i$--band 
magnitude selection with $17.5 \leq i_{AB} \leq 24$ and 
$23 \leq i_{AB} \leq 24.75$, respectively.
These two VVDS surveys are located in the VVDS--02h field centred at 
$\alpha_{2000} = 02 h 26 m 00 s$ 
and $\delta_{2000} = -04 \deg 30 \arcmin 00 \arcsec$. 

The VUDS is an on--going spectroscopic survey also using VIMOS, targeting $z>2$ 
galaxies in 3 fields:  COSMOS, ECDFS and VVDS--02h. 
The baseline target selection for spectroscopy is using photometric--redshifts 
$z_{phot}>2.4$ measured using all the photometry available in the survey
fields using the Le Phare code (Ilbert et al. 2006). 
The photometric redshift accuracy obtained from the multi--wavelength data and 
calibrated  on existing spectroscopic redshifts is  
$\sigma_{\delta z/(1+z)} \simeq 0.01$ for magnitudes $i_{AB}<25$
in the COSMOS field (see e.g. Ilbert et al. 2013).
There is a slight degradation by a factor $<2$ on the photometric
redshift accuracy in the other two fields because of the lesser number of 
photometric bands observed. 
In addition, we supplemented the 
$z_{phot}$ selection by several color--selection criteria, adding
those galaxies likely to be in this redshift range, but not selected 
from the primary $z_{phot}$ selection. 
Allowing for errors in $z_{phot}$, this selection provides a sample with 
$2 \lesssim z \lesssim 6$, as described  in Le F\`evre et al. 
(2013b, in preparation).

All VIMOS data have been processed with the VIPGI package 
(Scodeggio et al. 2005). 
Following automated  measurements, each galaxy has been examined visually and 
independently by two persons, each assigning a spectroscopic redshift.
These measurements have been compared before assigning the final redshift 
measurement. A reliability flag has been assigned to each redshift measurement 
representing the probability for the redshift to be right. As consistently
shown from the VVDS (Le F\`evre et al. 2005), zCOSMOS (Lilly et al. 2007),
and VIPERS (Guzzo et al. 2013) surveys, the reliability of flags
reflect the statistical process of redshift assignment between independant
observers and does not depend on
the survey type or its intrinsic quality, with flags 1, 2, 3, 4 
and 9 having a probability to be right of $\sim 50, 87, 98, 100, 90\%$, 
respectively  (flags 1x with x=1, 2, 3, 4, 9 indicate a broad line AGN; flag 2x are objects 
falling serendipitously 
in a slit next to a main target; and both have probability 
distributions similar to the main flag categories).

In addition to the VIMOS spectroscopic data, a large set of imaging data is 
available in the three fields covered by our pair search. 
The COSMOS field (Scoville et al. 2007) has a full coverage with the HST/ACS 
F814W filter (Koekemoer et al. 2007) and includes, among other data, ugriz 
photometry from Subaru (Taniguchi et al. 2007), and more recently YJHK 
photometry from the UltraVista survey (McCracken et al. 2012). 
Spectroscopic redshifts from the zCOSMOS survey are also available 
(Lilly et al. 2007). 
The ECDFS is covered by the MUSYC survey in UBVRIz' (Gawiser et al. 2006) and 
partly by the CANDELS survey with the ACS and WFC3 on HST 
(Koekemoer et al. 2011).
The VVDS--02h field has deep CFH12K BVRI photometry (Le F\`evre et al. 2004),
and even deeper CFHTLS ugriz photometry (e.g. Ilbert et al. 2006),
as well as JHK photometry from the deep survey with CFHT-WIRCAM (WIRDS, Bielby et al. 2012).


\section{Pair identification}
\label{pair_select}

Pairs have been identified using the projected transverse separation $r_p$ and 
the velocity difference $\Delta v = c ( v_1 - v_2 ) / (1-v_1v_2)$ where $v_x$, the
normalized velocity of the $x^{th}$ galaxy, is given as
$v_x = [(1+z_x)^2 - 1] / [(1+z_x)^2 + 1]$. 
The survey samples have first been scanned for separation 
$r_p \leq 25$h$^{-1}$kpc and $\Delta v \leq 500$ km s$^{-1}$. 
We chose these separations as it is expected that pairs would merge in about 1 Gyr 
(e.g. Kitzbichler \& White, 2008), meaning that a pair observed at $z \sim 3$ 
would have merged by the peak in star formation activity at $z\sim1.5-2$ 
(e.g. Cucciati et al. 2012). 

Given the limitation of our ground based seeing observations, we are not able 
to identify pairs superimposed along the line of sight or separated spatially
by less than 1 arcsec, the average image quality (FWHM) of the imaging
data, corresponding to 5 h$^{-1}$kpc at $z\sim3$.
In addition, pairs with larger $r_p$ or $\Delta$v could also merge, albeit on a 
longer timescale, and with a lower probability (Kitzbichler \& White, 2008). 
We will expand our pair sample to these larger separations and will correct for 
the geometrical limitations and selection function of the observations, 
following e.g. de Ravel et al. (2009),
when addressing the merger rate in a forthcoming paper. 

We used all objects with reliability flags 2 to 9 (galaxies) and 12 
to 19 (broad--line AGN) for the primary targets, including a total of 1111 
galaxies,
and an additional 811 galaxies with flags 1 to 9 and 11 to 19, as well as 21 to 29 (objets 
falling serendipitously in the slit), for the companions. 

In looking for pairs, there is a possibility that the two objects identified
could be 
two giant HII regions of the same galaxy, which would then be prominent in the 
UV rest--frame probed by the $i$--band, but which would appear as 
a single galaxy in the $H$ or $K$ band which probes redward of the D4000 or 
Balmer breaks at wavelengths less sensitive
to contamination by younger stellar populations. On the contrary,
the persistence of separate morphological components in the redder bands is an 
indication of the existence of two galaxies even if these are embedded in a 
diffuse light background.  
We therefore examined the $i$--band and $H/K$--band images of the pair 
candidates selected from their separation.
As described in the next section, all of our pairs are well separated
and both galaxies are seen from bluer to redder bands.
We are
therefore confident that we are dealing with true physical pairs 
rather than double HII regions of the same galaxy. 

In this process, we have identified 12 pairs with redshifts from $z=1.82$ to 
$z=3.65$. All primary and companion
galaxies have flags$\geq2$ except one galaxy 
(ID:520478087) with a lower reliability flag 1. Our pair sample therefore
benefits from reliable spectroscopic identification.
The pair properties are discussed in the next section.


\section{Pair properties}
\label{pairs_prop}

\begin{figure*}
   \centering
   \includegraphics[width=4.7cm]{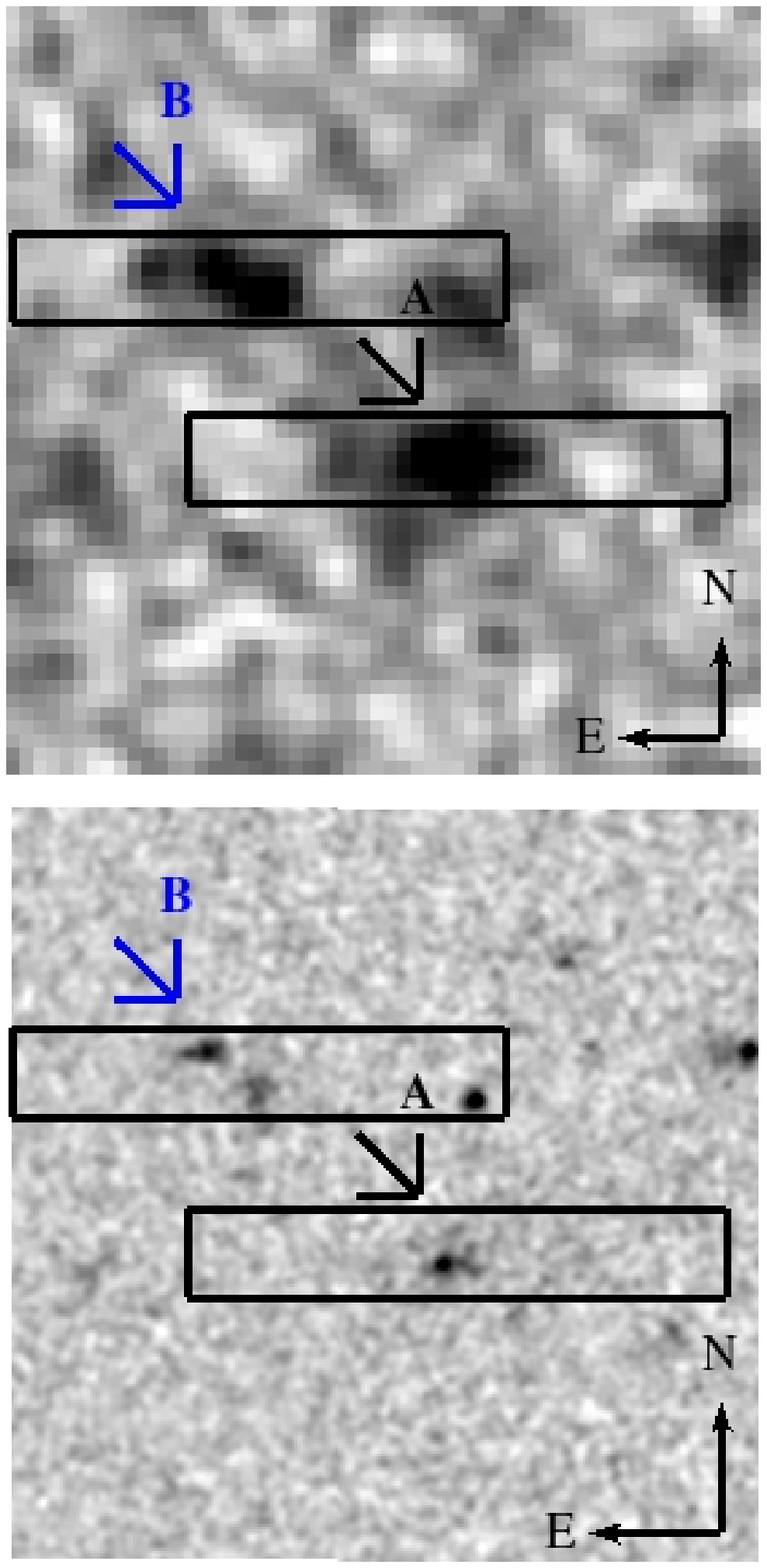}
   \includegraphics[width=9cm]{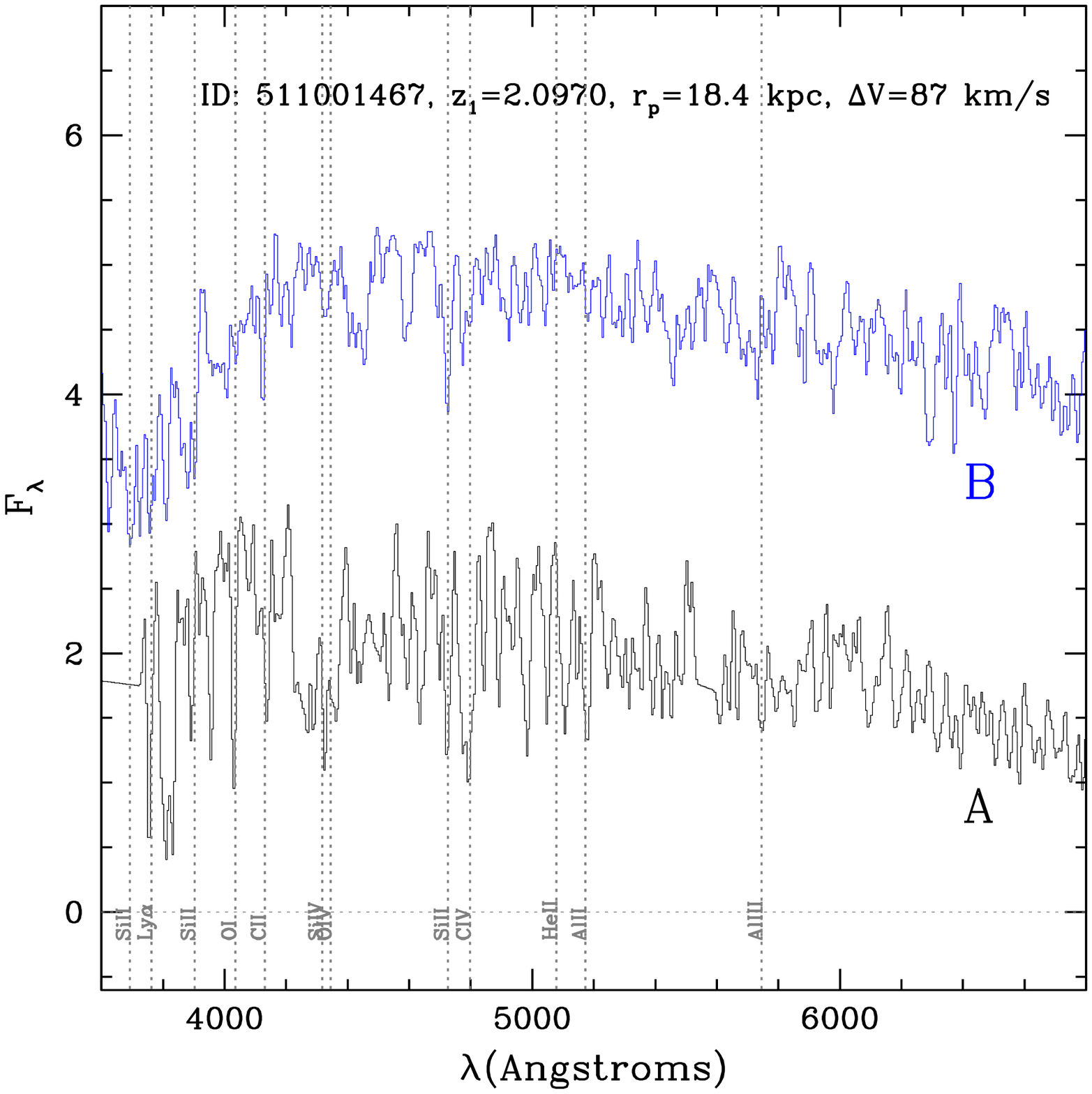}
      \caption{Pair COSMOS--511001467 A/B: $5\arcsec \times 5$\arcsec sum of 
      UltraVista YJHK images (top--left) and HST F814W image (bottom--left).
      The location of the 1 arcsec width VLT/VIMOS slits is shown by the rectangles.
      The more massive object is labeled as A.
      Right Panel: VIMOS spectra for both components in the pair.
      The spectra have been arbitrarly shifted in flux to avoid overlap.}
         \label{pairs_1}
   \end{figure*}

\begin{figure*}
   \centering
   \includegraphics[width=4.5cm]{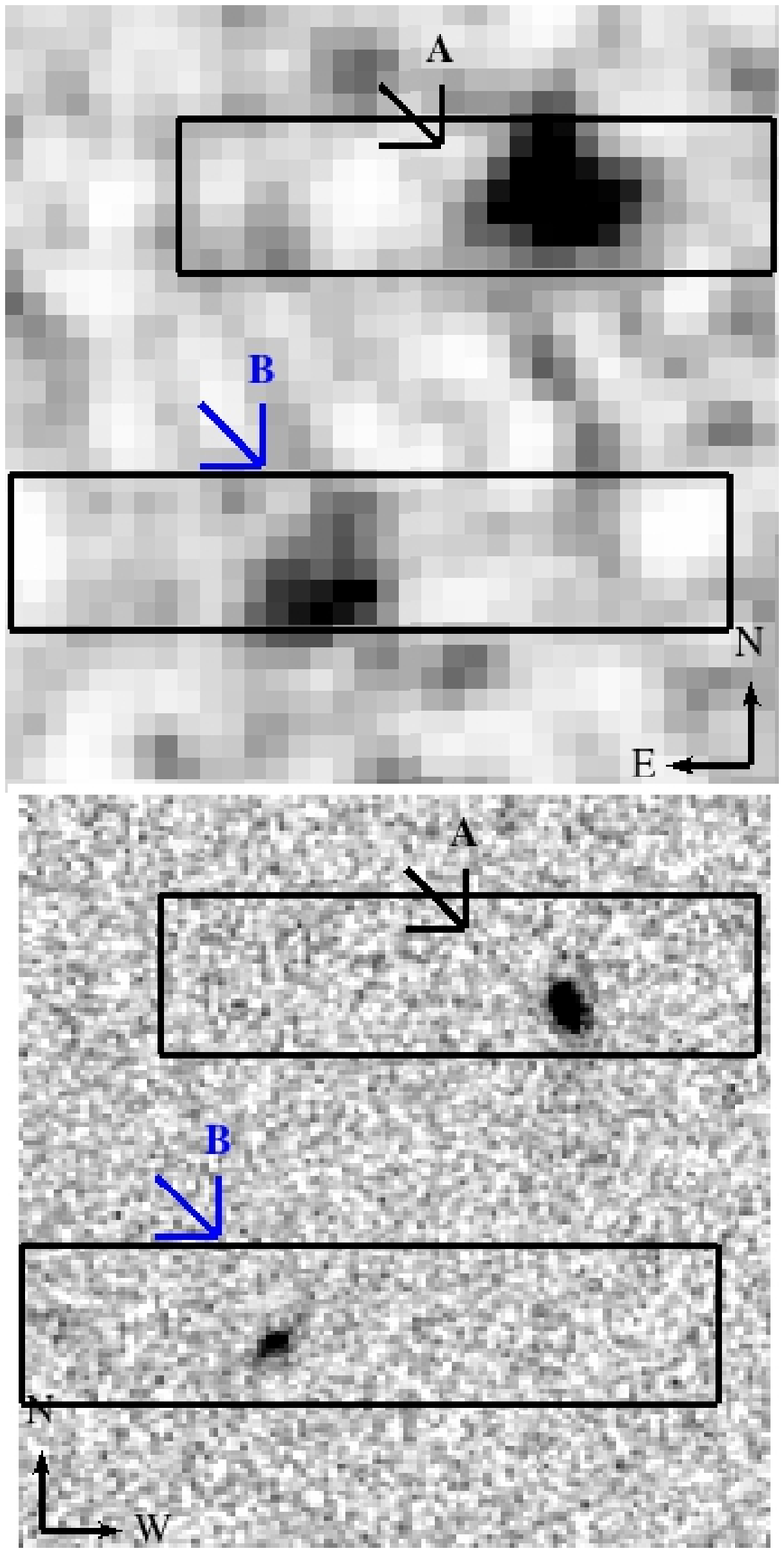}
   \includegraphics[width=9cm]{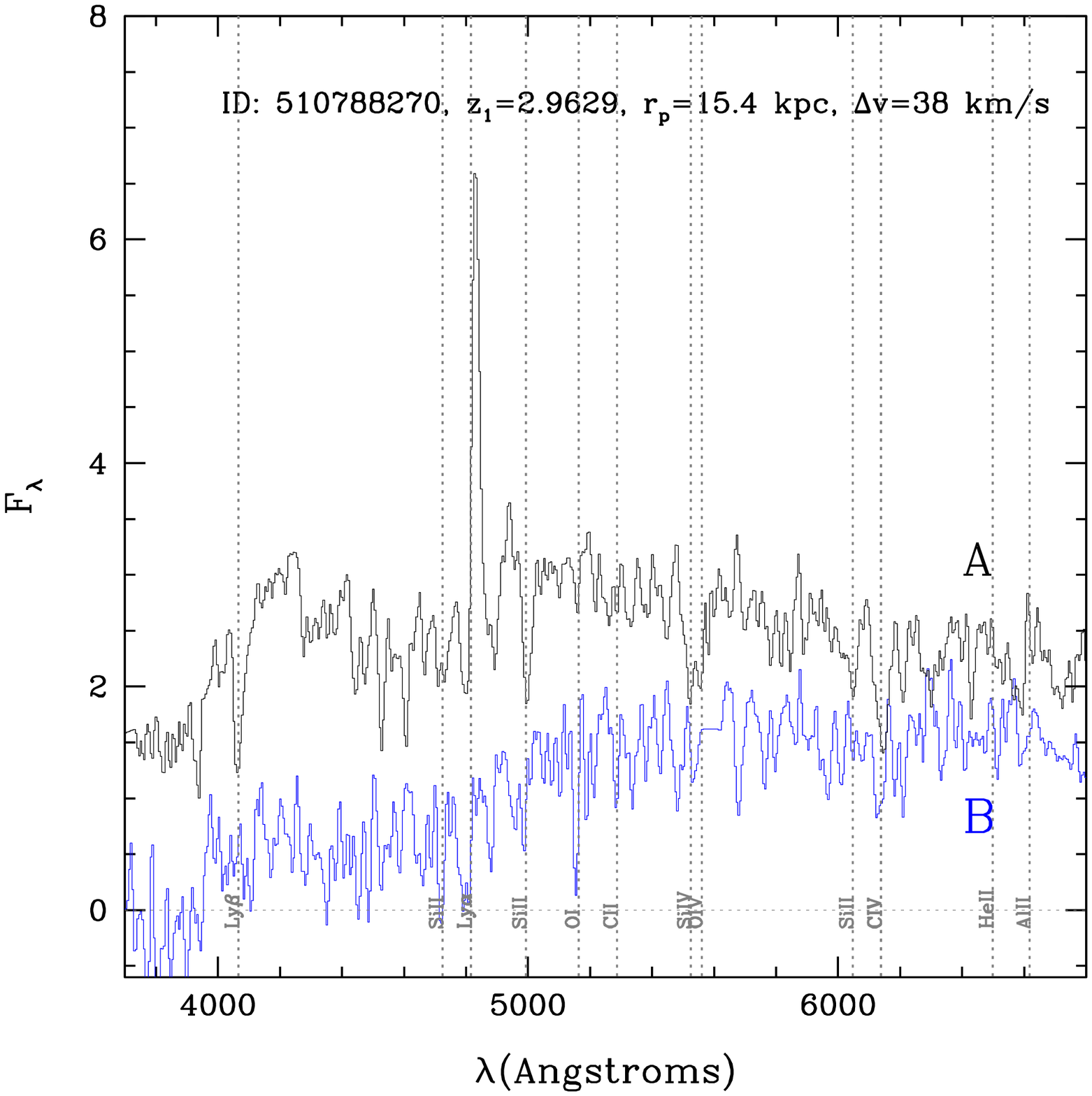}
      \caption{Pair COSMOS--510788270 A/B: $5\arcsec \times 5$\arcsec sum of 
      UltraVista YJHK images (top--left) and HST F814W image (bottom--left).
      The location of the 1 arcsec width VLT/VIMOS slits is shown by the rectangles.
      The more massive object is labeled as A.
      Right Panel: VIMOS spectra for both components in the pair.
      The spectra have been arbitrarly shifted in flux to avoid overlap.}
         \label{pairs_2}
   \end{figure*}

\begin{figure*}
   \centering
   \includegraphics[width=4.5cm]{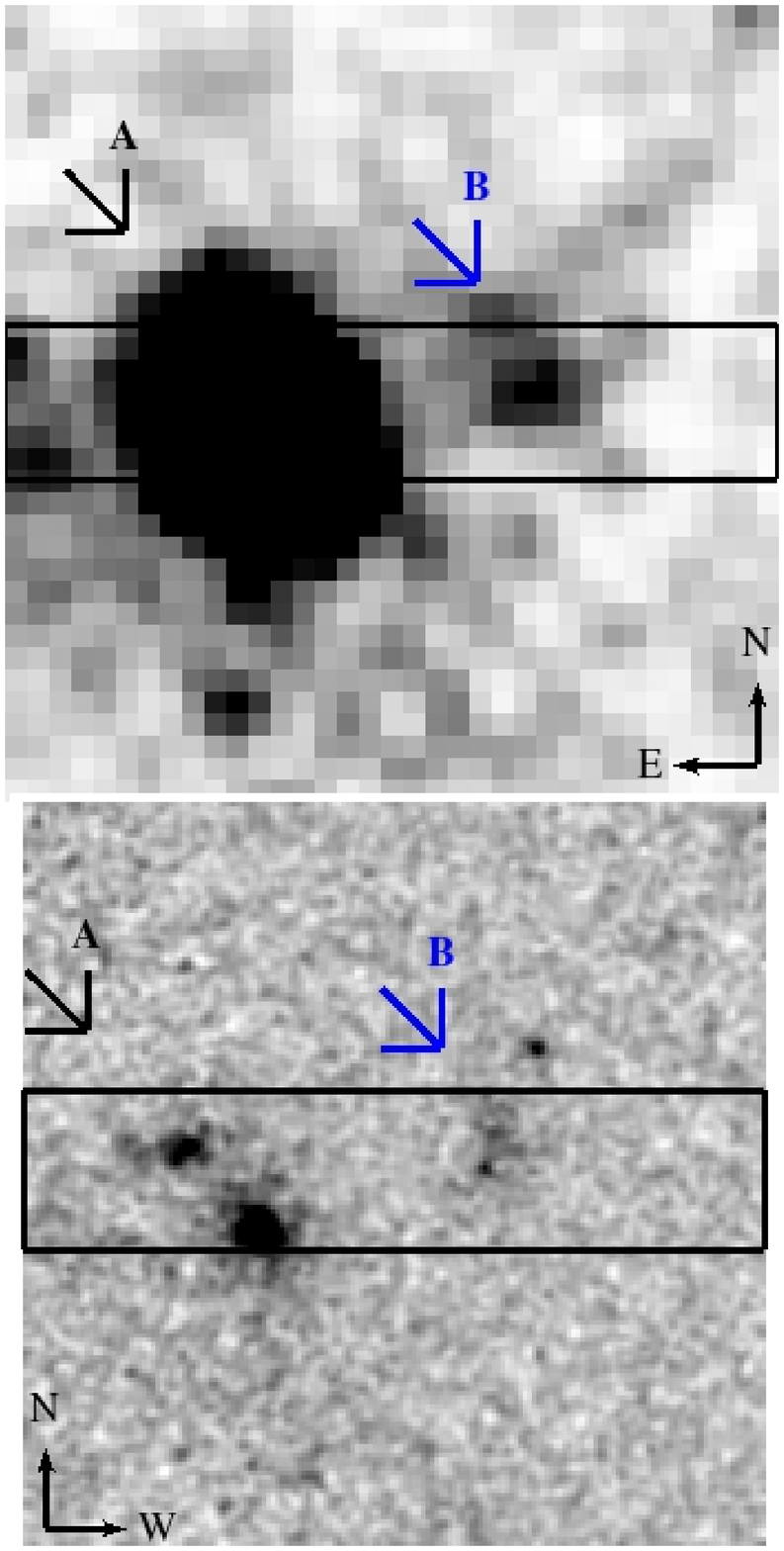}
   \includegraphics[width=9cm]{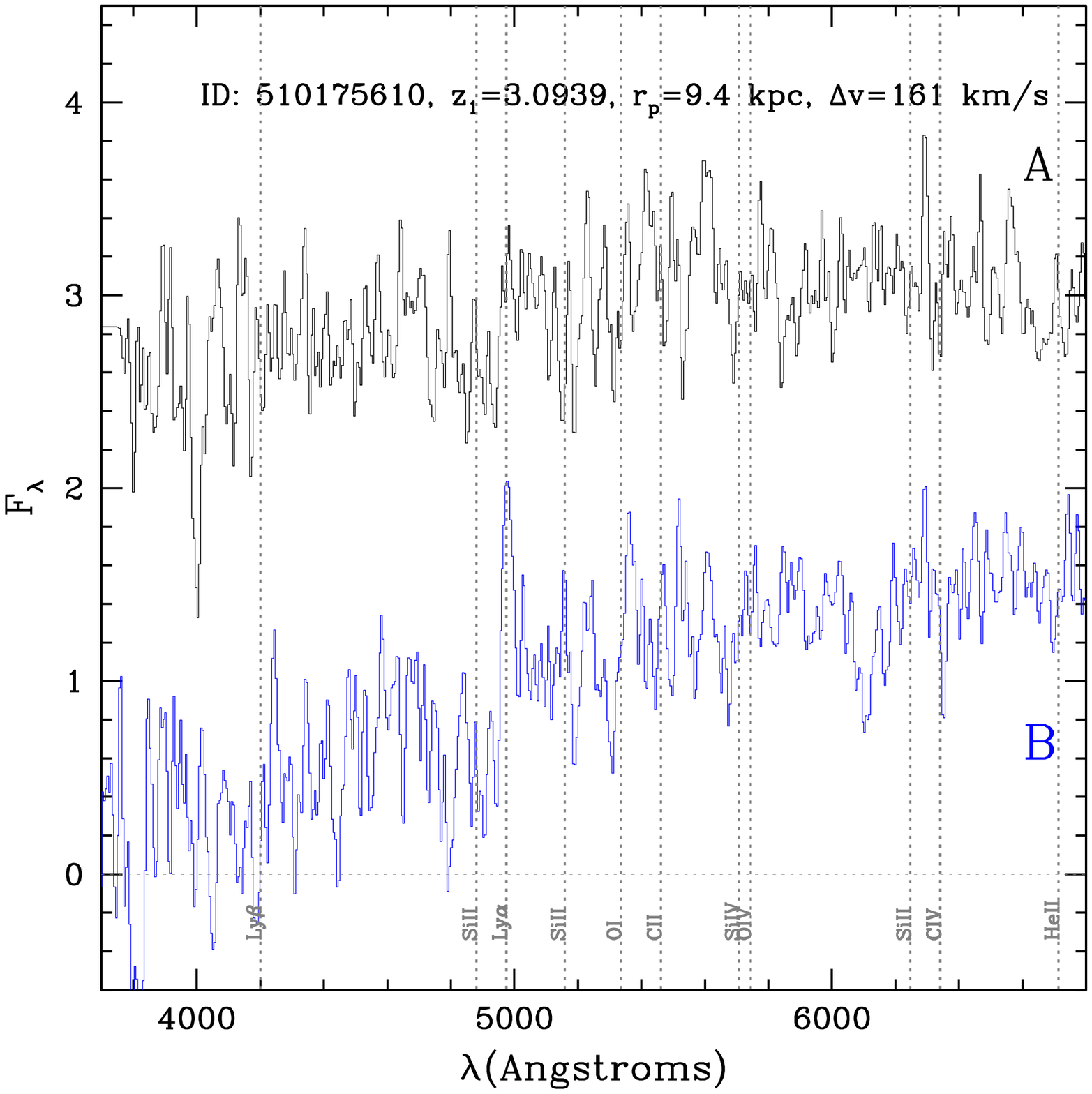}
      \caption{Pair COSMOS--510175610/510778438: $5\arcsec \times 5$\arcsec sum 
      of UltraVista YJHK images (top--left) and HST F814W image (bottom--left).
      The location of the 1 arcsec width VLT/VIMOS slit is shown by the rectangle.
      The more massive object is labeled as A.
      Right Panel: VIMOS spectra for both components in the pair.
      The spectra have been arbitrarly shifted in flux to avoid overlap.
      }
         \label{pairs_3}
   \end{figure*}

\begin{figure*}
   \centering
   \includegraphics[width=4.5cm]{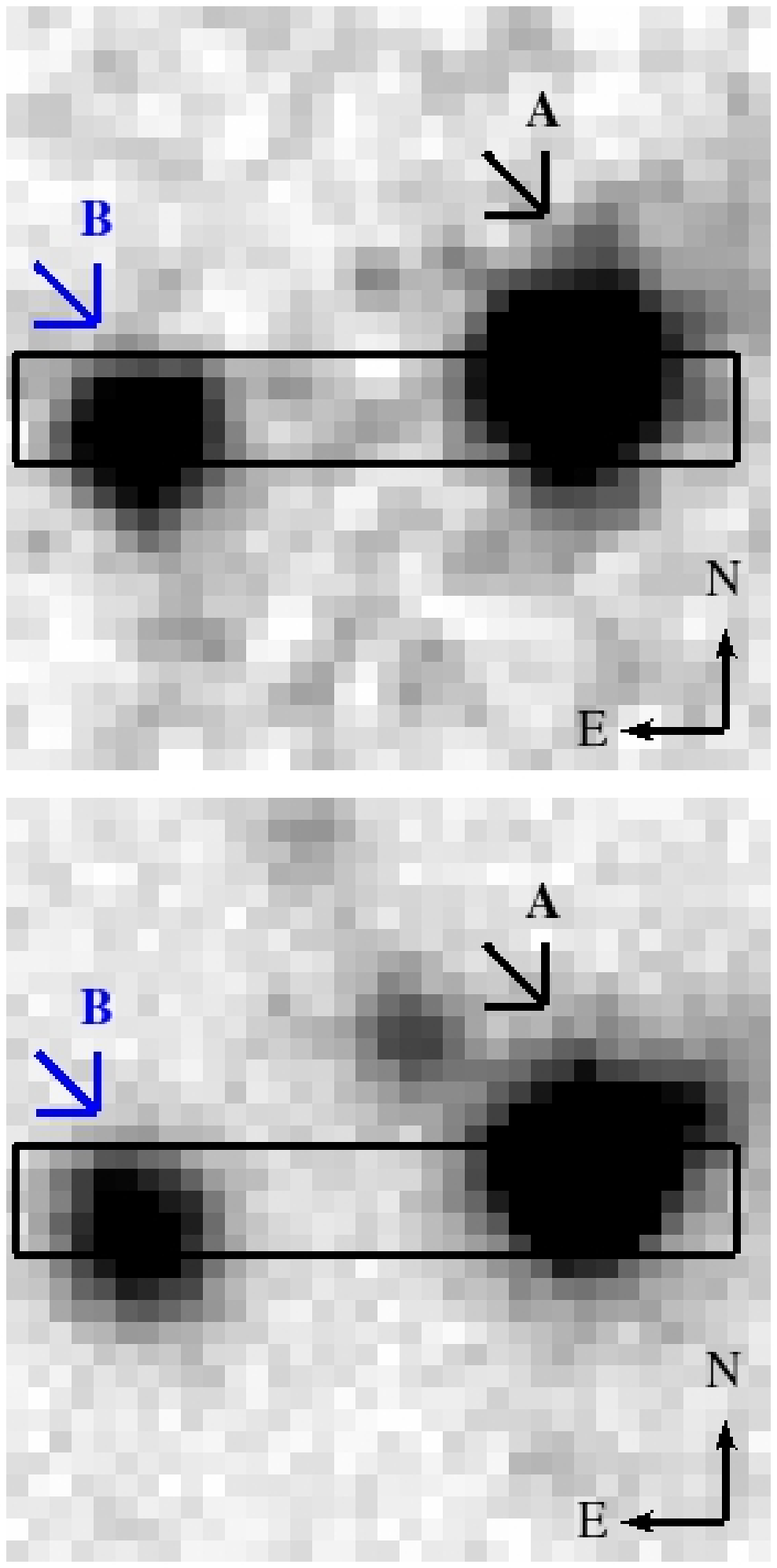}
   \includegraphics[width=9cm]{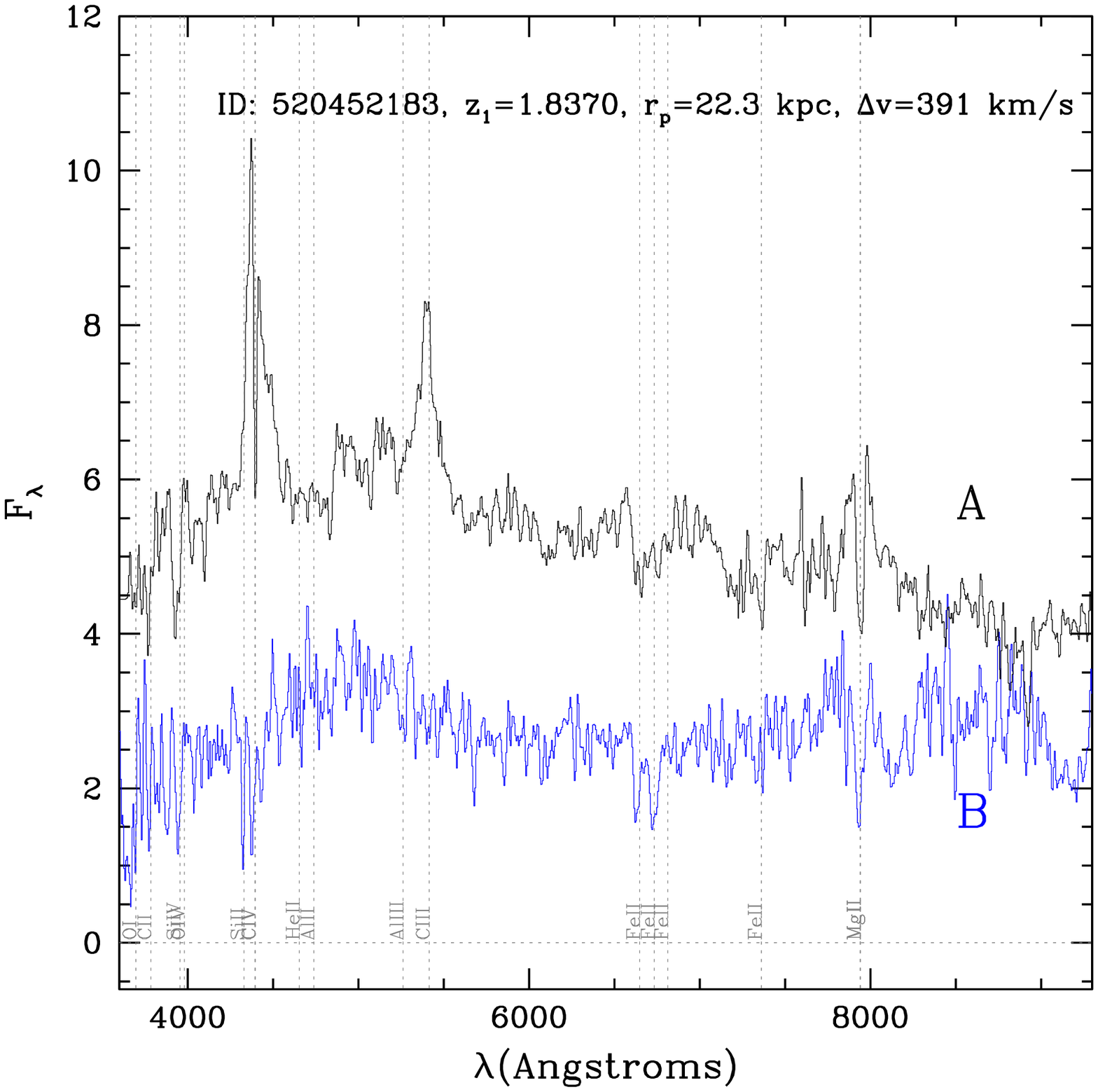}
      \caption{Pair VVDS--2h--520452183/520450423: 
      $6.5\arcsec \times 6.5$\arcsec composite WIRDS JHKs image (top--left) and
      CFHTLS i--band image (bottom--left). 
      The location of the 1 arcsec width VLT/VIMOS slit is shown by the rectangle.
      The more massive object is labeled as A.
      Right panel: VIMOS spectra for both components in the pair.
      The spectra have been arbitrarly shifted in flux to avoid overlap. 
              }
         \label{pairs_4}
   \end{figure*}

\begin{figure*}
   \centering
   \includegraphics[width=4.5cm]{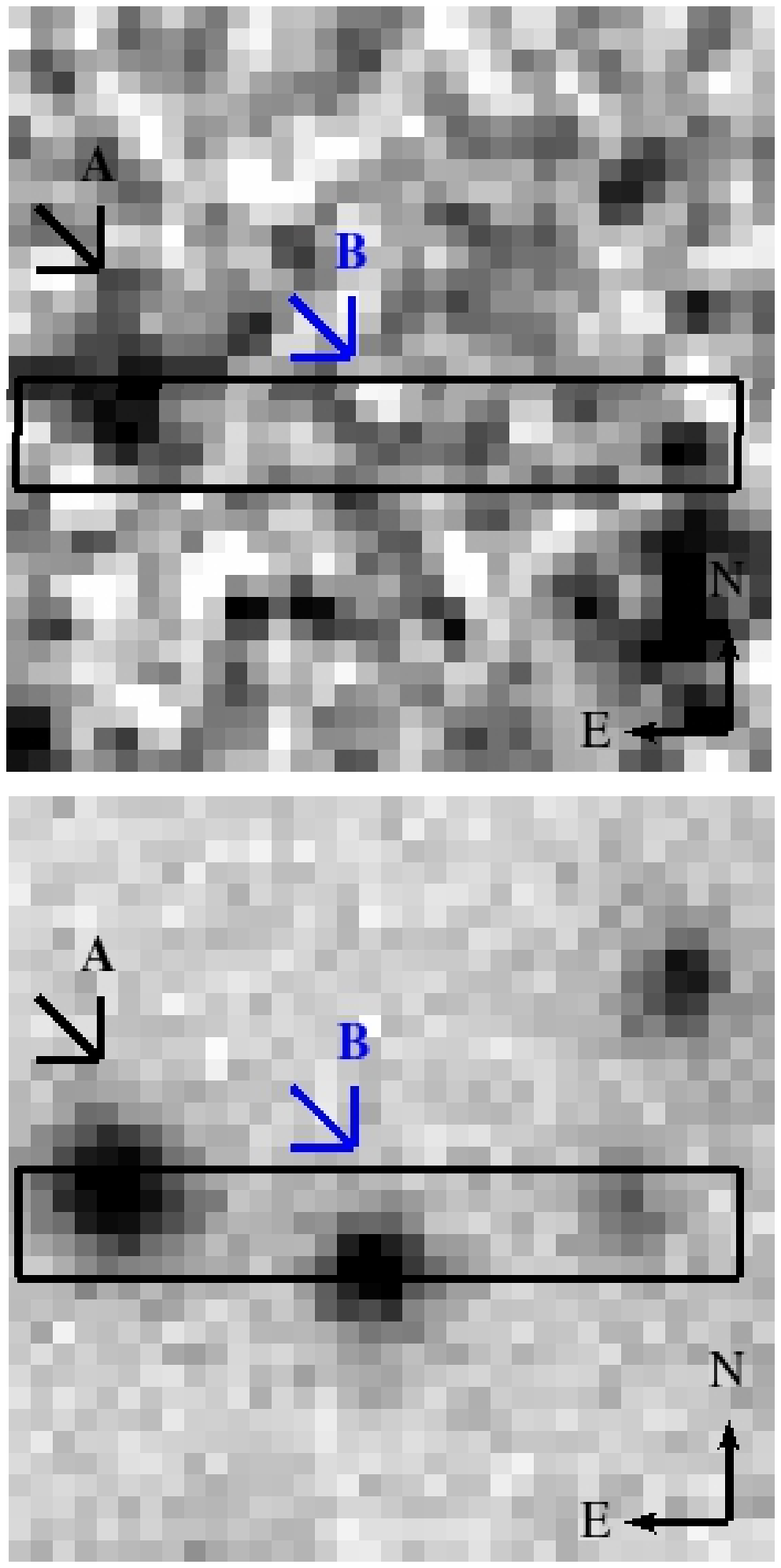}
   \includegraphics[width=9cm]{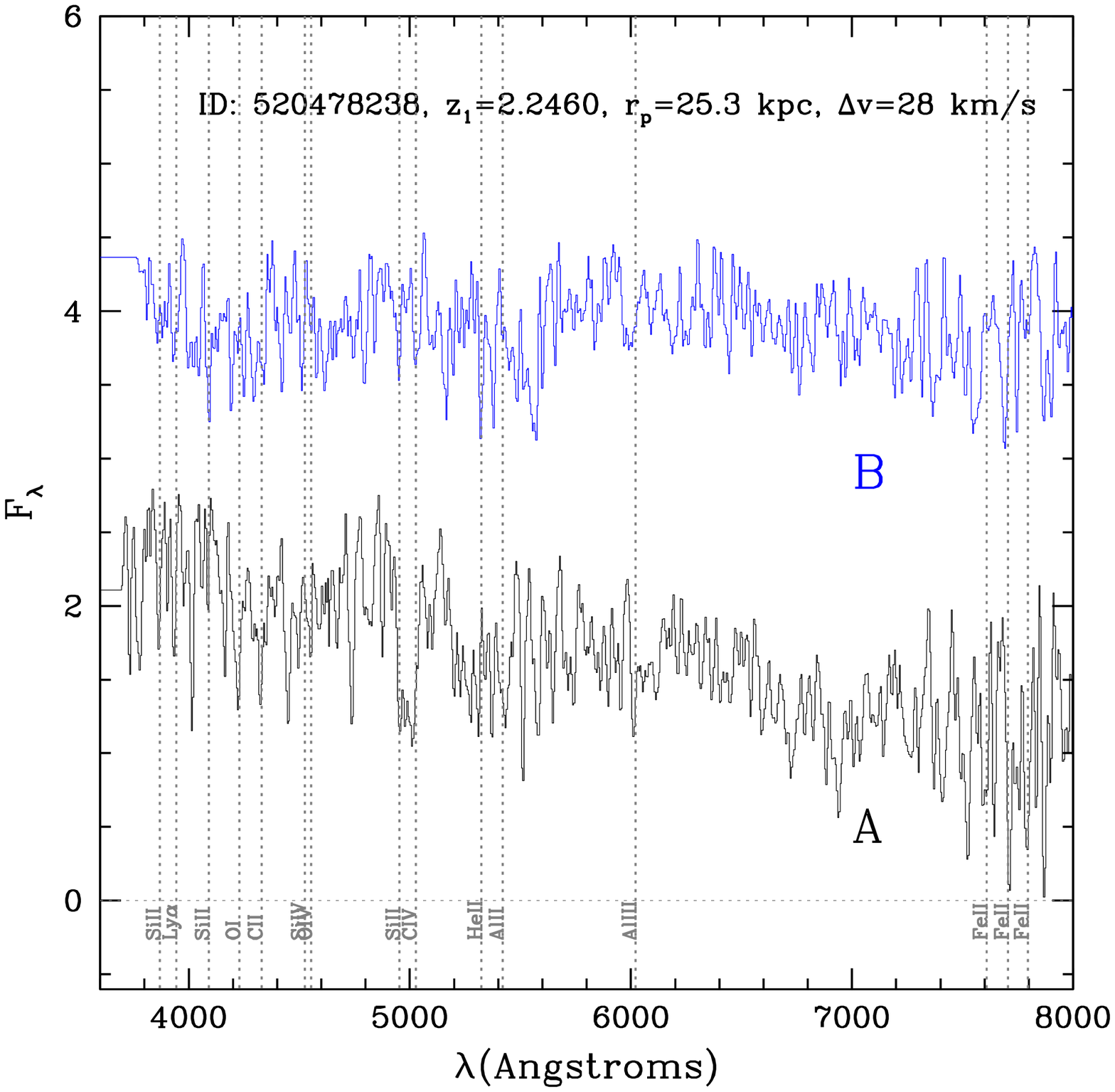}
      \caption{Pair VVDS--2h--520478238/520478087: 
      $6.5\arcsec \times 6.5$\arcsec composite WIRDS JHKs image (top--left) and
      CFHTLS i--band image (bottom--left). 
      The location of the 1 arcsec width VLT/VIMOS slit is shown by the rectangle.
      The more massive object is labeled as A.
      Right panel: VIMOS spectra for both components in the pair. 
      The spectra have been arbitrarly shifted in flux to avoid overlap.
              }
         \label{pairs_5}
   \end{figure*}

\begin{figure*}
   \centering
   \includegraphics[width=4.5cm]{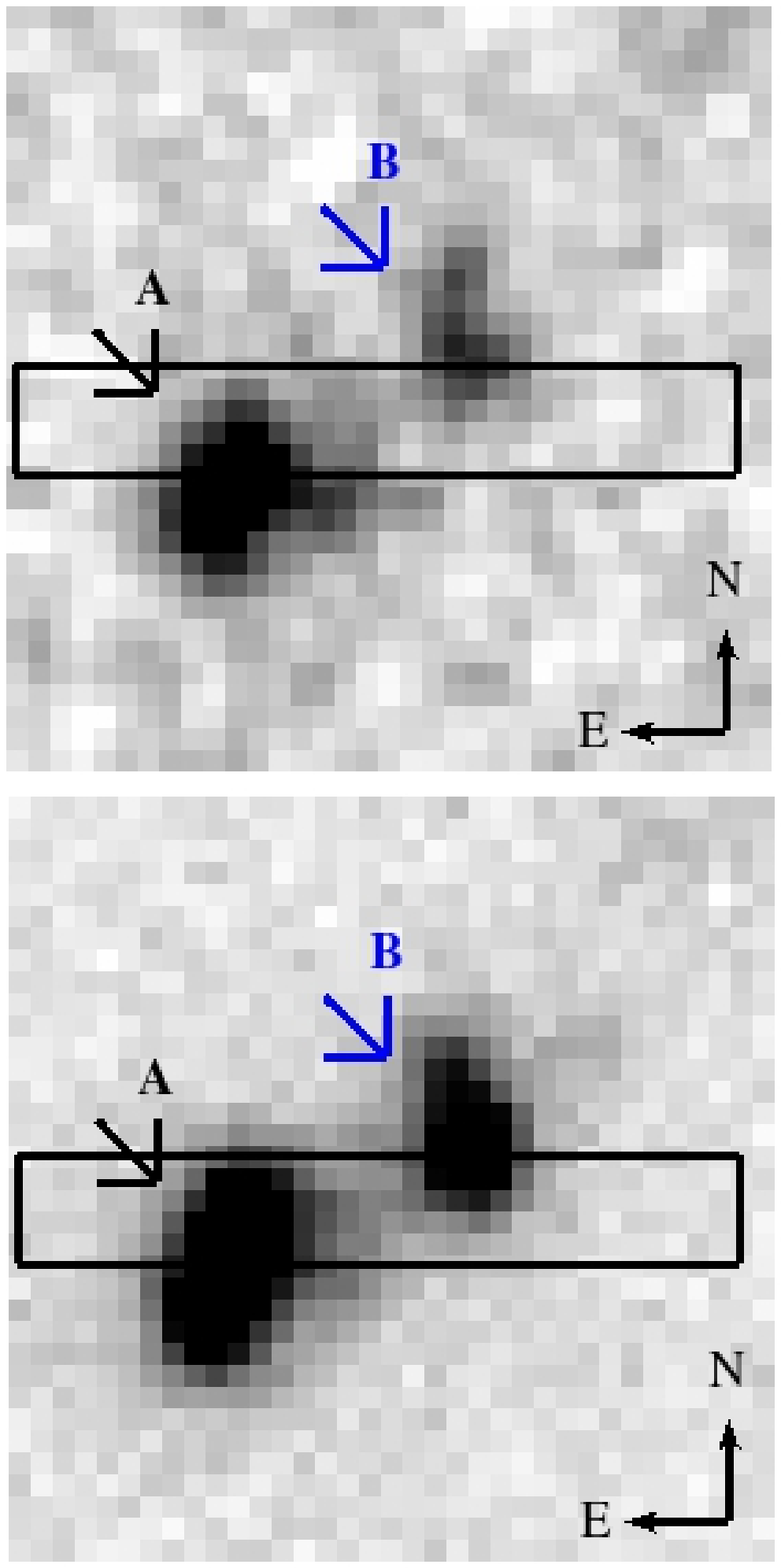}
   \includegraphics[width=9cm]{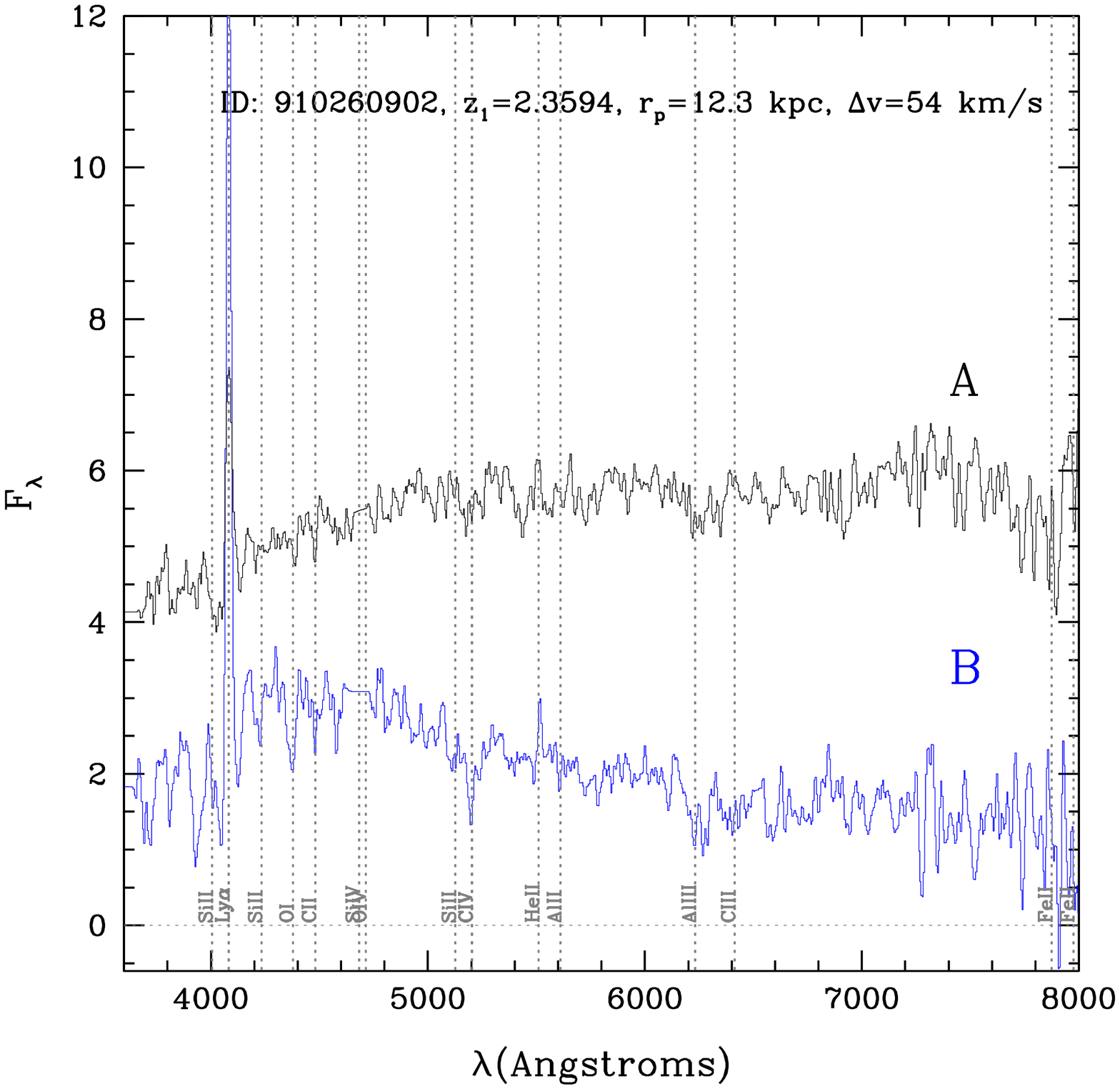}
      \caption{Pair VVDS--2h--910260902/910261083: 
      $6.5\arcsec \times 6.5$\arcsec composite WIRDS JHKs image (top--left) and
      CFHTLS i--band image (bottom--left). 
      The location of the 1 arcsec width VLT/VIMOS slit is shown by the rectangle.
      The more massive object is labeled as A.
      Right panel: VIMOS spectra for both components in the pair. 
      The spectra have been arbitrarly shifted in flux to avoid overlap.       
              }
         \label{pairs_6}
   \end{figure*}

\begin{figure*}
   \centering
   \includegraphics[width=4.5cm]{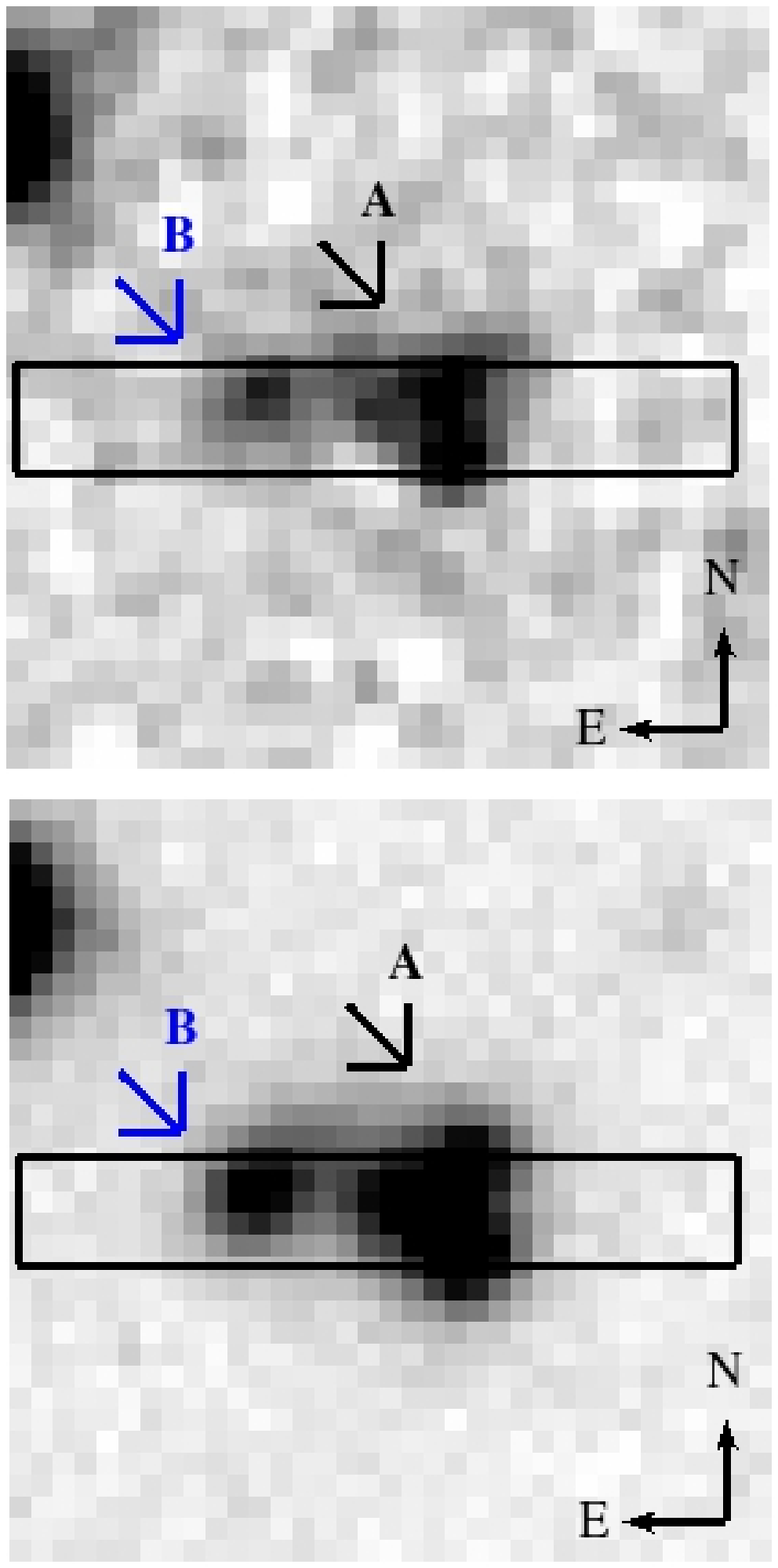}
   \includegraphics[width=9cm]{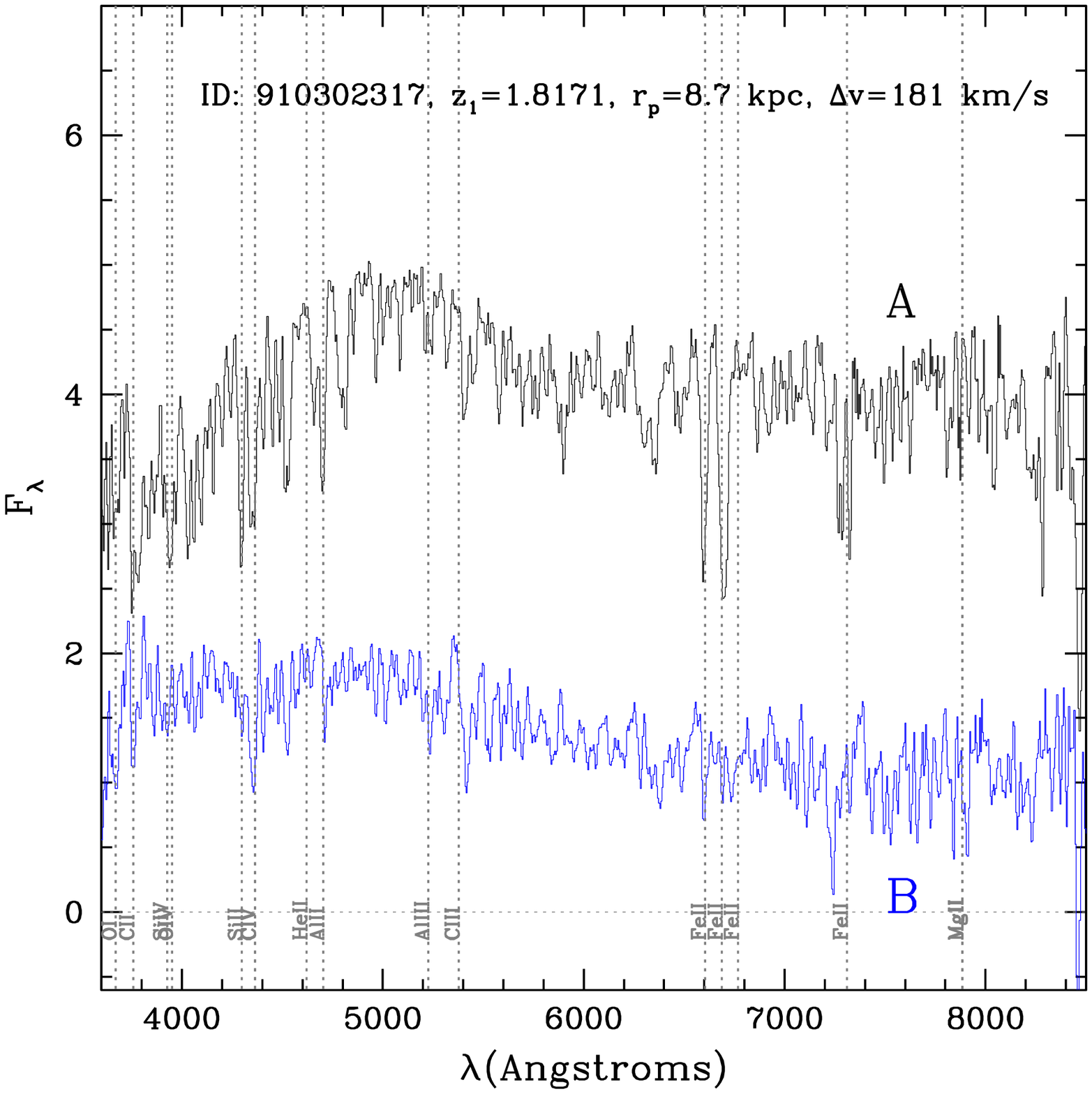}
      \caption{Pair VVDS--2h--910302317/910302929: 
      $6.5\arcsec \times 6.5$\arcsec composite WIRDS JHKs image (top--left) and
      CFHTLS i--band image (bottom--left). 
      The location of the 1 arcsec width VLT/VIMOS slit is shown by the rectangle.
      The more massive object is labeled as A.
      Right panel: VIMOS spectra for both components in the pair.
      The spectra have been arbitrarly shifted in flux to avoid overlap.        
              }
         \label{pairs_7}
   \end{figure*}

\begin{figure*}
   \centering
   \includegraphics[width=4.5cm]{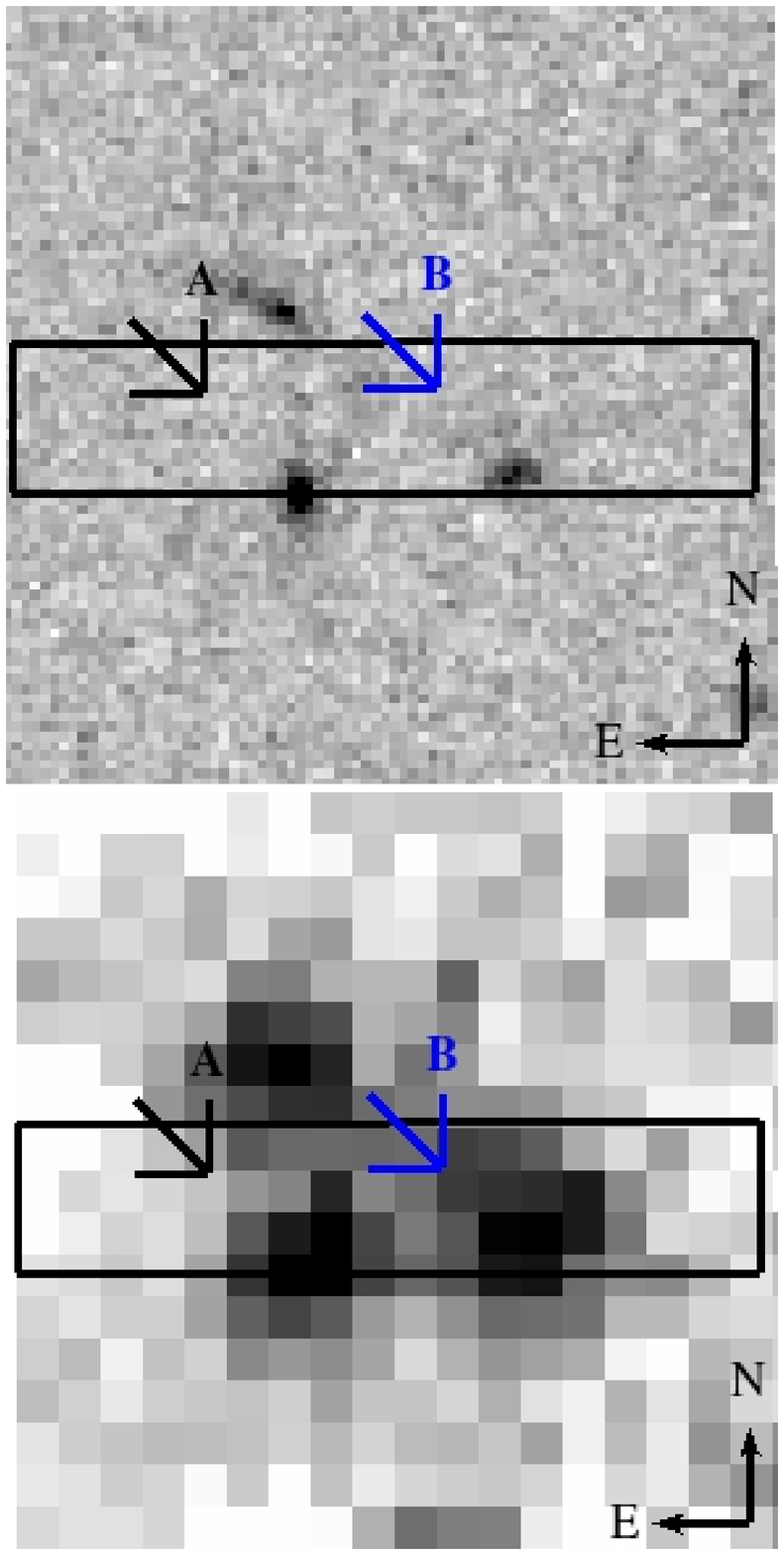}
   \includegraphics[width=9cm]{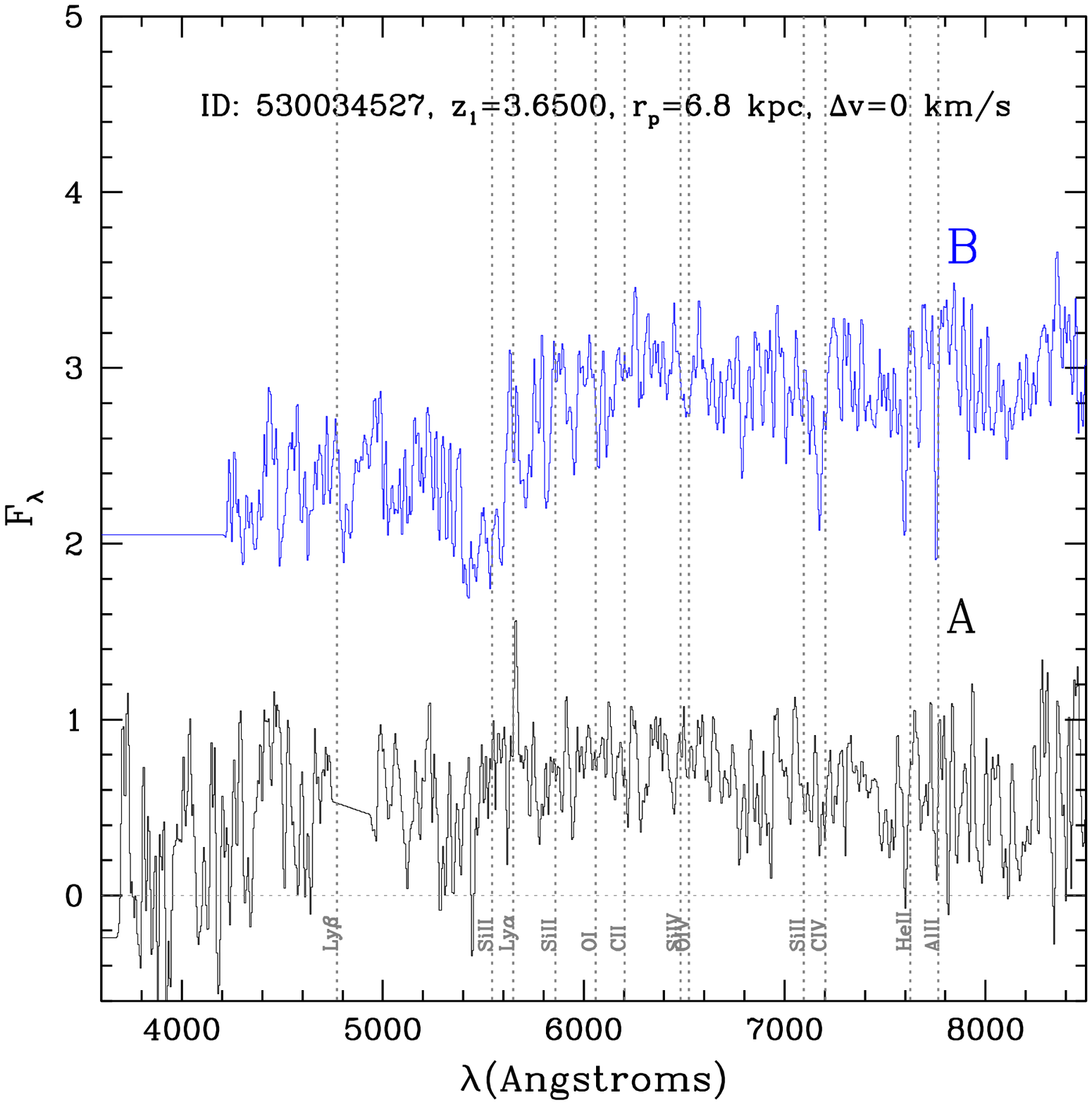}
      \caption{Pair ECDFS--530034527 A/B: $5\arcsec \times 5$\arcsec HST/WFC3 
      F160W CANDELS image (top--left) and composite BVR image from the MUSYC 
      survey MUSYC (bottom--left). 
      The location of the 1 arcsec width VLT/VIMOS slit is shown by the rectangle.
      The more massive object is labeled as A.
      Right panel: VIMOS spectra for both components in the pair.
      The spectra have been arbitrarly shifted in flux to avoid overlap. 
              }
         \label{pairs_8}
   \end{figure*}

\begin{figure*}
   \centering
   \includegraphics[width=4.5cm]{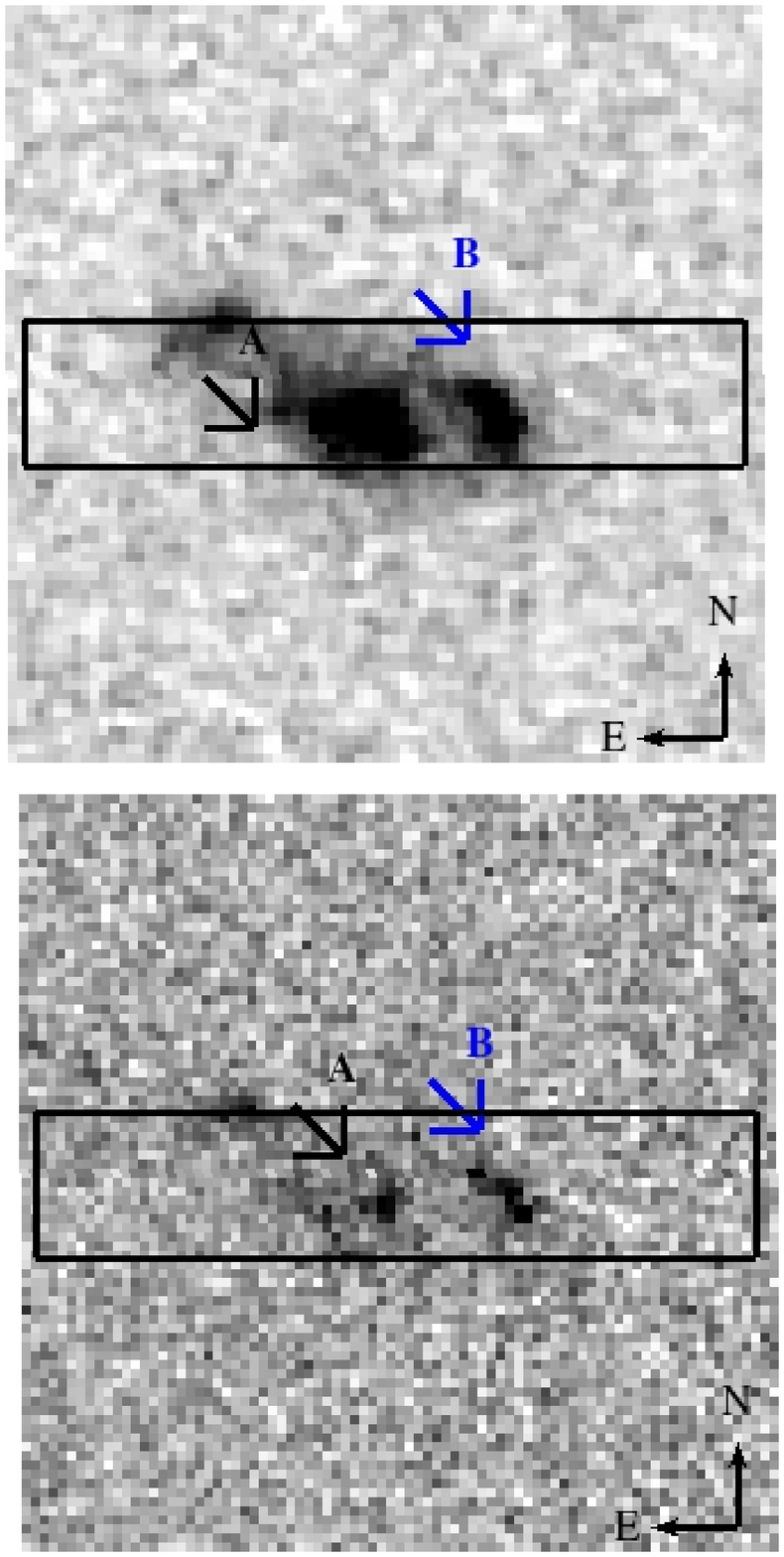}
   \includegraphics[width=9cm]{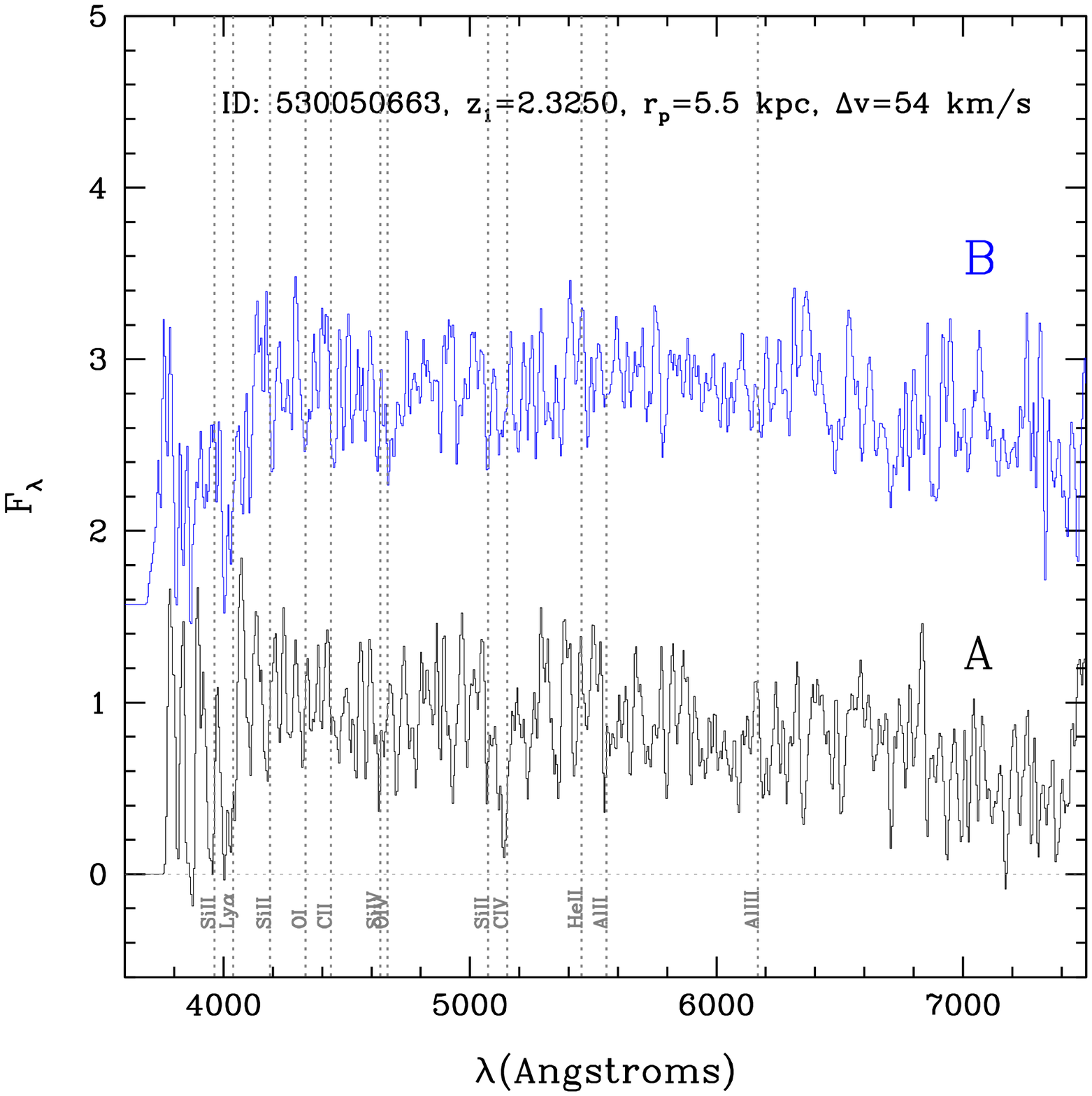}
      \caption{Pair ECDFS--530050663 A/B: $5\arcsec \times 5$\arcsec HST/WFC3 
      F160W image (top--left) and HST/ACS F850W image from the CANDELS survey 
      (bottom--left).
      The location of the 1 arcsec width VLT/VIMOS slit is shown by the rectangle.
      The more massive object is labeled as A.
      Right panel: VIMOS spectra for both components in the pair. 
      The spectra have been arbitrarly shifted in flux to avoid overlap.
              }
         \label{pairs_9}
   \end{figure*}

\begin{figure*}
   \centering
   \includegraphics[width=4.5cm]{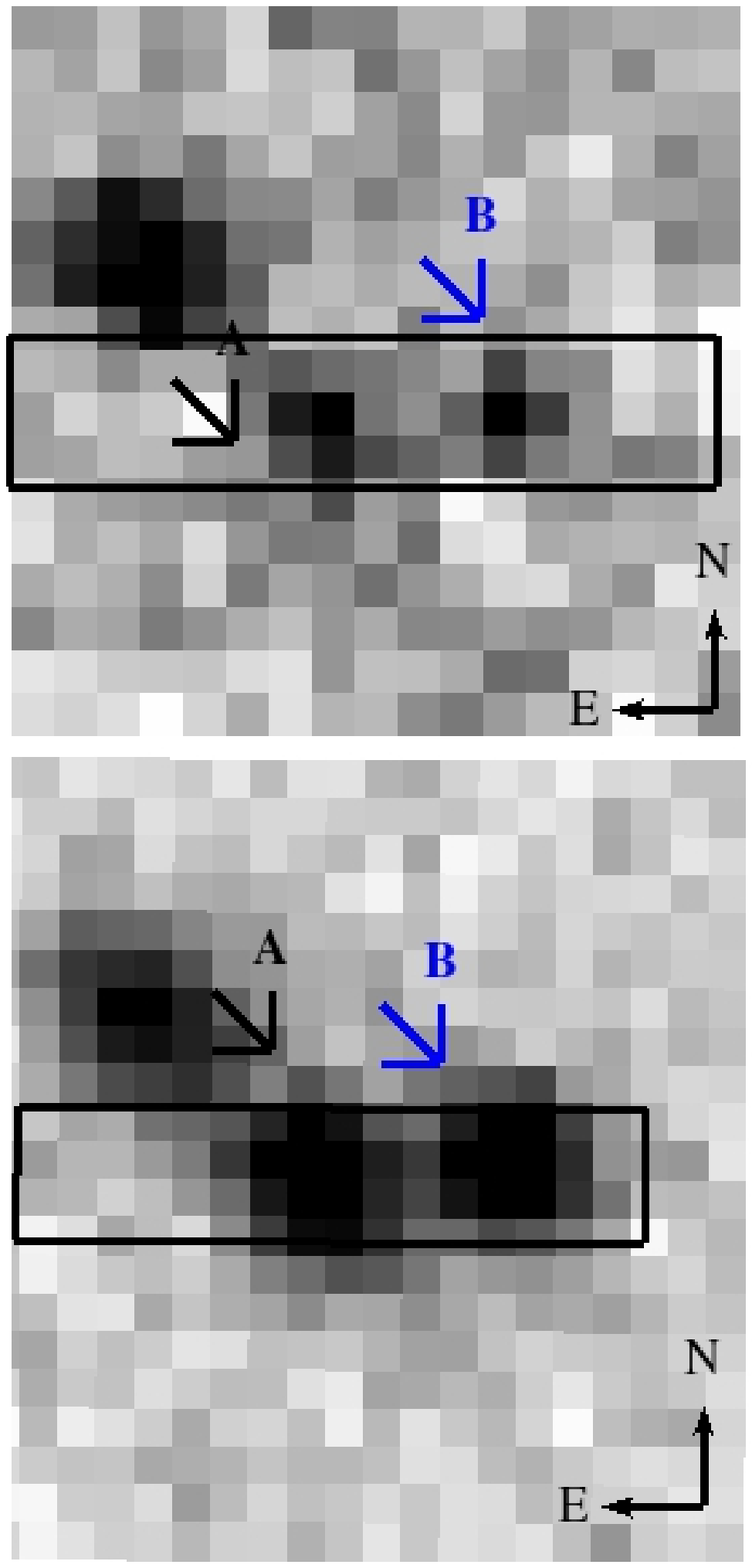}
   \includegraphics[width=9cm]{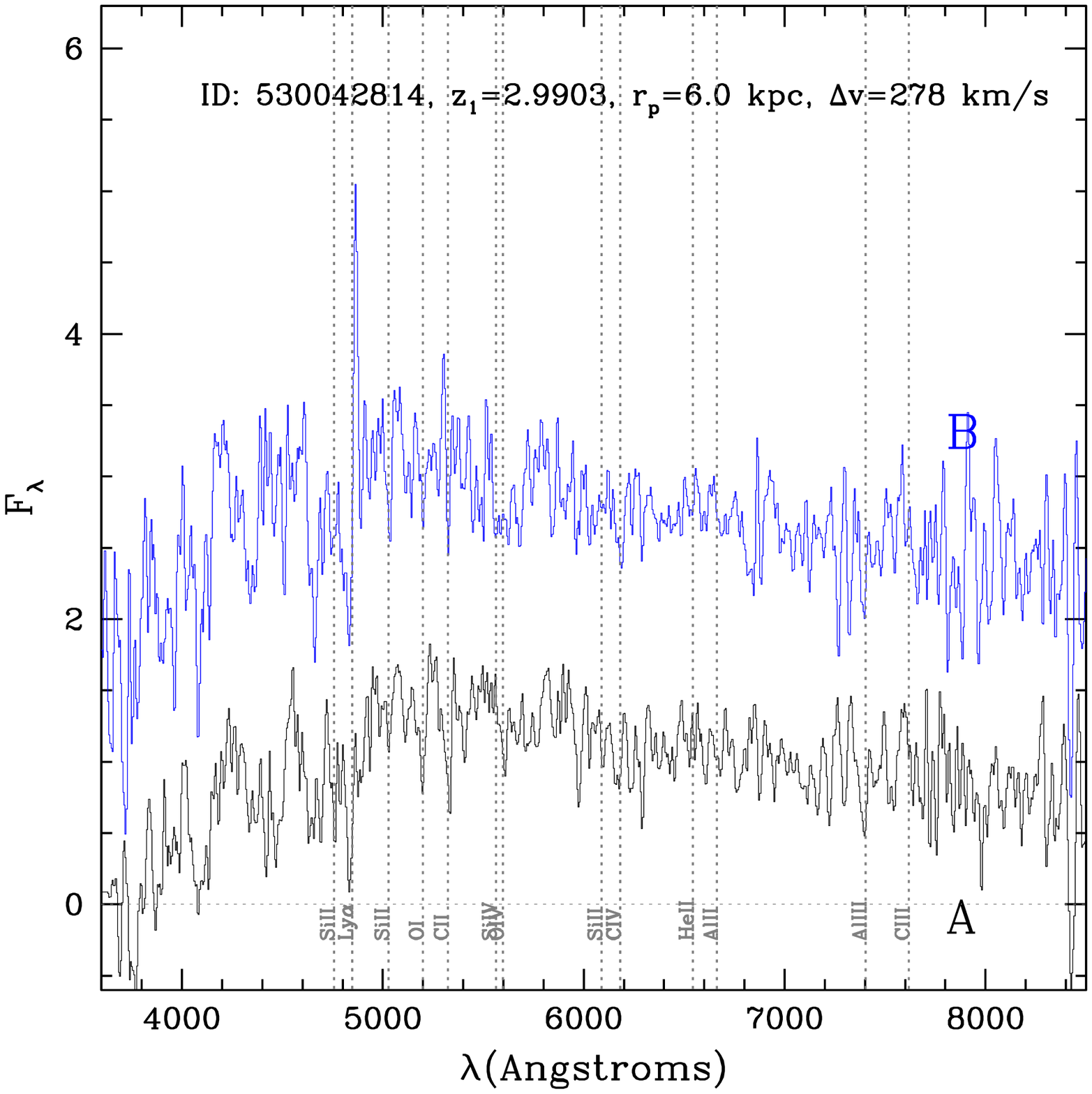}
      \caption{Pair ECDFS--530042814/530042840: $5\arcsec \times 5$\arcsec 
      composite JK image 
      (top--left) and composite BVR image from the MUSYC survey (bottom-left).
      The location of the 1 arcsec width VLT/VIMOS slit is shown by the rectangle.
      The more massive object is labeled as A.
      Right panel: VIMOS spectra for both components in the pair. 
      The spectra have been arbitrarly shifted in flux to avoid overlap.
              }
         \label{pairs_10}
   \end{figure*}

\begin{figure*}
   \centering
   \includegraphics[width=4.5cm]{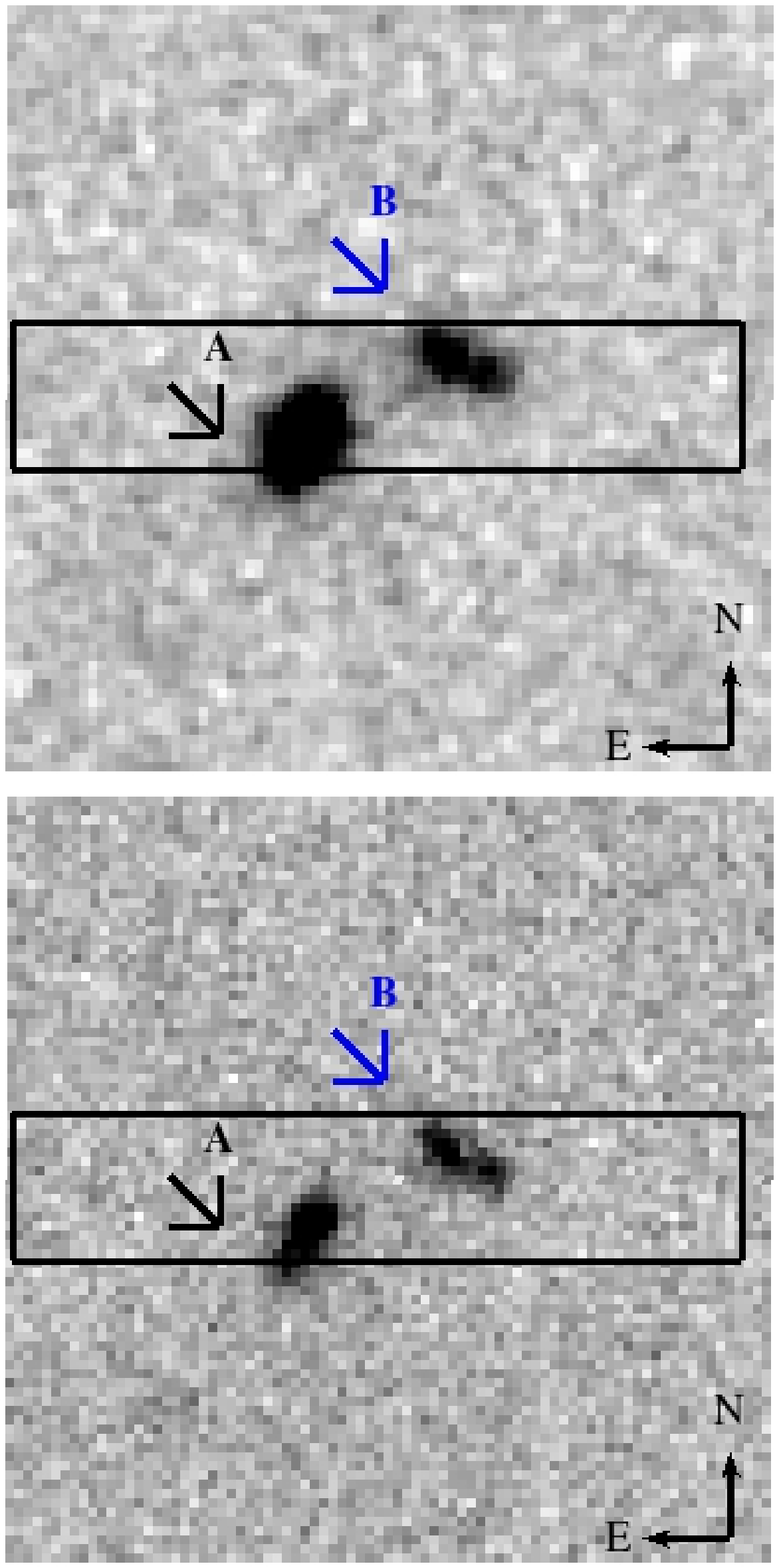}
   \includegraphics[width=9cm]{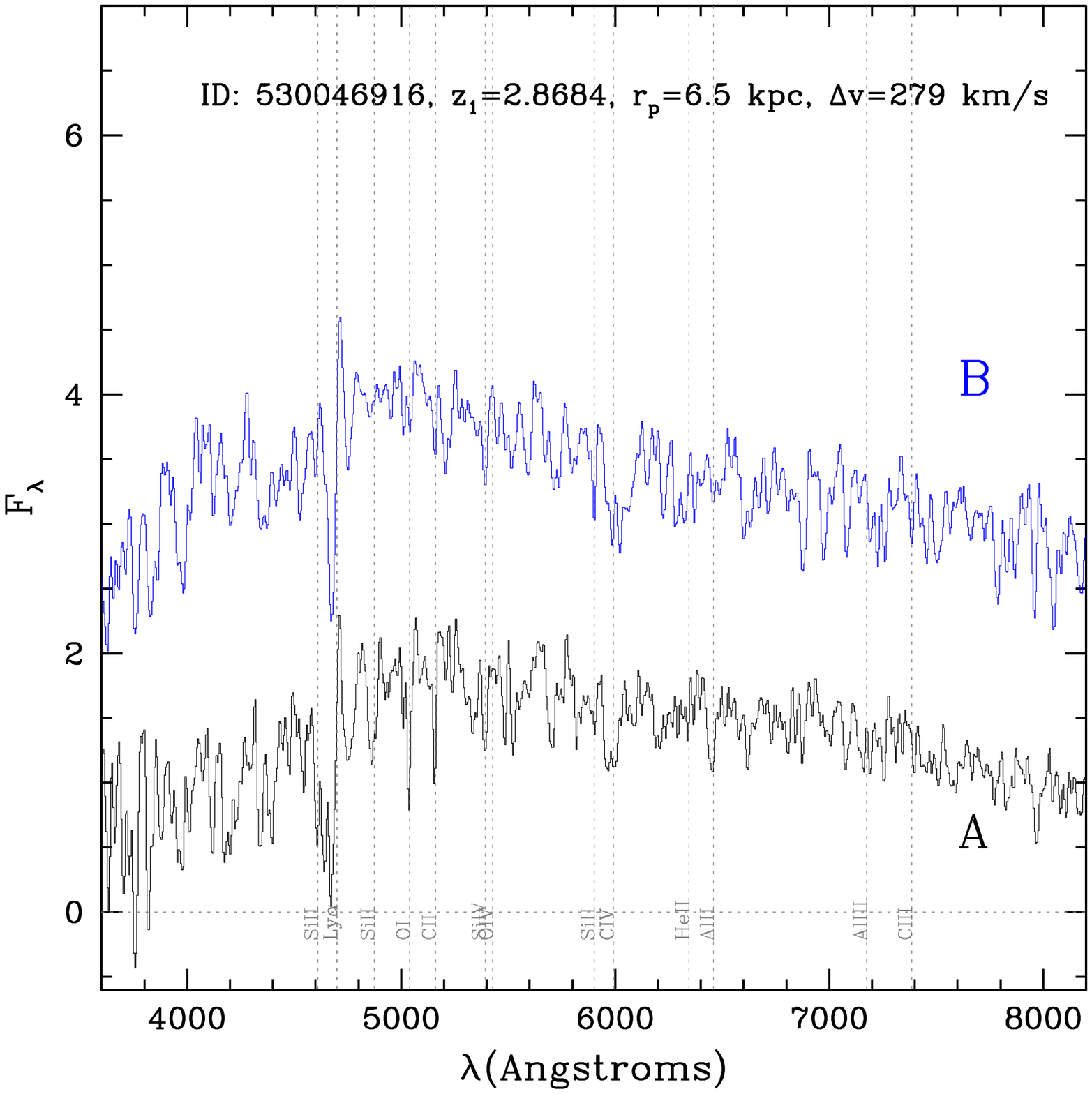}
      \caption{Pair ECDFS--530046916 A/B: $5\arcsec \times 5$\arcsec HST/WFC3 
      F160W image (top--left) and HST/ACS F850W image from the CANDELS survey 
      (bottom--left). 
      The location of the 1 arcsec width VLT/VIMOS slit is shown by the rectangle.
      The more massive object is labeled as A.
      Right panel: VIMOS spectra for both components in the pair. 
      The spectra have been arbitrarly shifted in flux to avoid overlap.
              }
         \label{pairs_11}
   \end{figure*}

\begin{figure*}
   \centering
   \includegraphics[width=4.5cm]{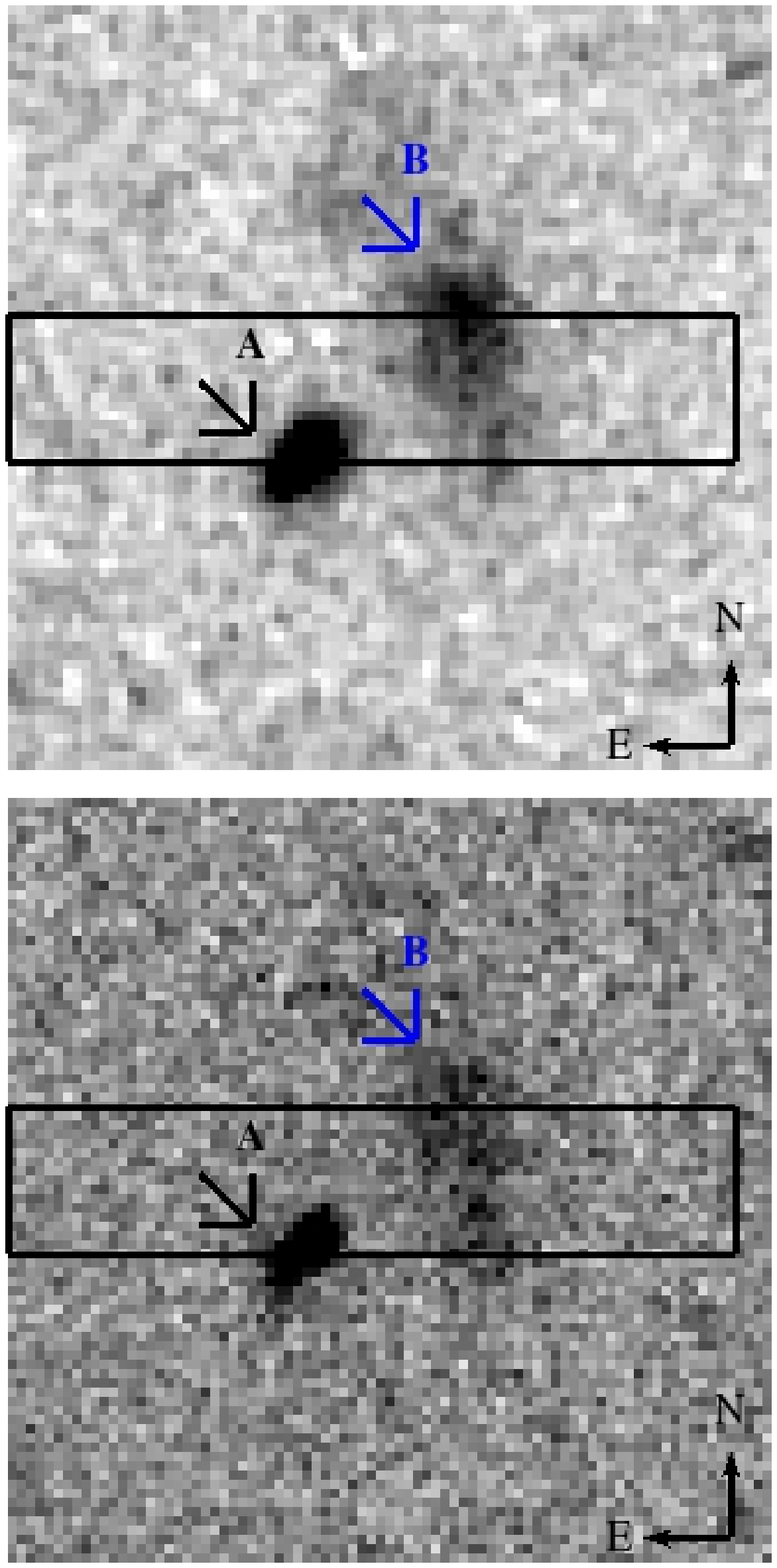}
   \includegraphics[width=9cm]{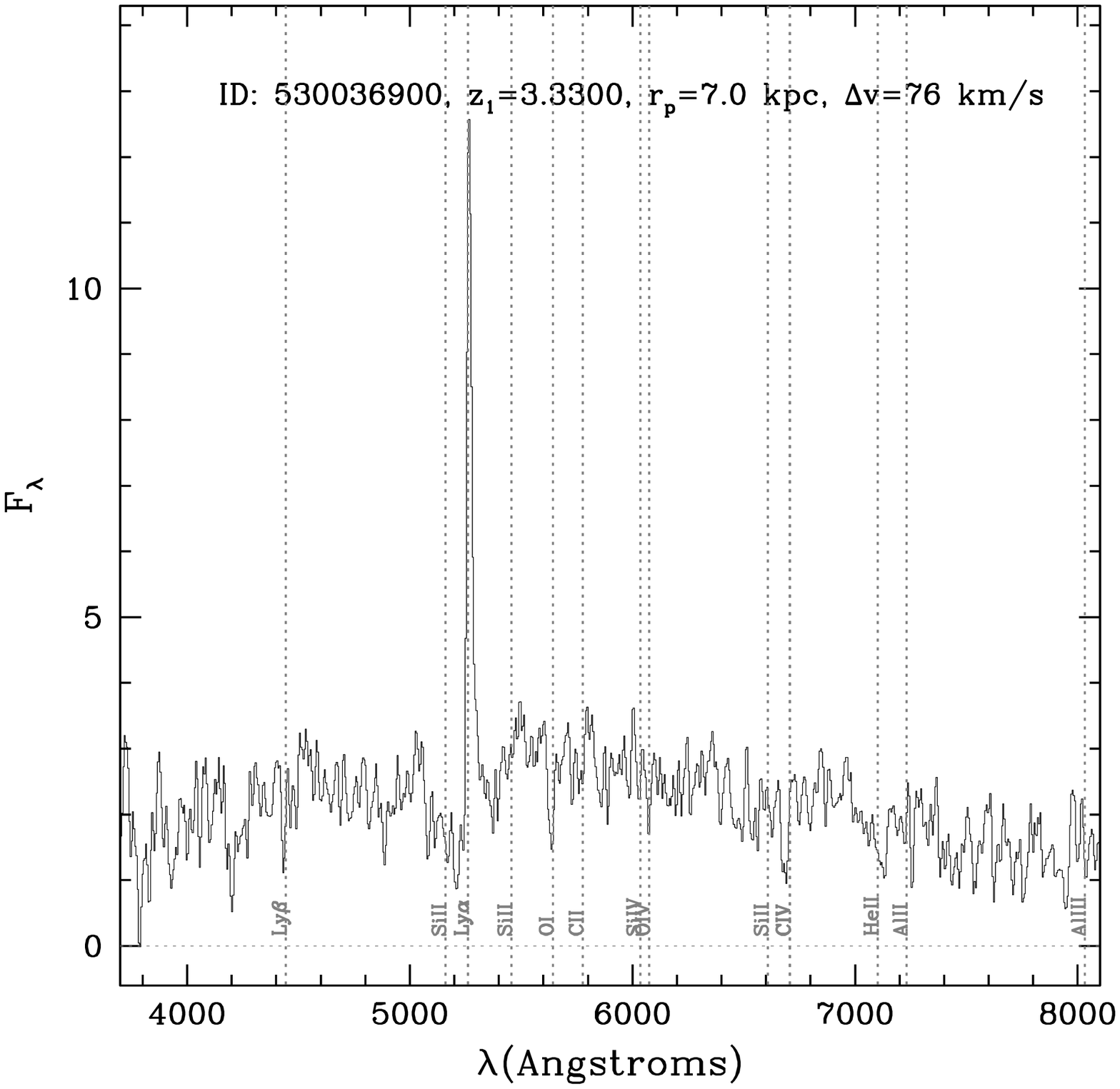}
      \caption{Pair ECDFS--530036900 A/B: $5\arcsec \times 5$\arcsec HST/WFC3 
      F160W image (top--left) and HST/ACS F850W image from the CANDELS survey 
      (bottom--left). 
      The location of the 1 arcsec width VLT/VIMOS slit is shown by the rectangle.
      The more massive object is labeled as A.
      Right panel: VIMOS VUDS spectrum for component A. The redshift of
      component B has been measured by Popesso et al. (2009) and not shown in the figure.
              }
         \label{pairs_12}
   \end{figure*}

From the multi--wavelength dataset, we have derived the main properties of the 
galaxies in the identified pairs. 
The stellar mass of each galaxy of a pair has been derived from SED fitting of the available photometric 
data at the  measured redshift (see e.g. Ilbert et al. 2010) and  range from $10^9$ to $10^{11}$ M$_{\sun}$ 
(Table~\ref{tab_pairs}), with uncertainies by up to a factor 2 (e.g. Bolzonella et al. 2009)
depending on the number of bands used, or the IMF. 
The stellar mass ratio are used to separate ``major--merging pairs'', 
with  a  mass ratio between the two galaxies of $1 \leq M_1/M_2 \leq 4$,
from ``minor--merging pairs'', for which $M_1/M_2 > 4$.
We have checked that the SED--derived mass ratio for each pair is consistent 
with the $K$--band or $H$--band flux ratio of the two galaxies, as these fluxes 
probing above $\lambda_{rest} \gtrsim 4000$\AA ~can be considered as a rough 
proxy for stellar mass (Bruzual and Charlot, 2003), and we have used the error 
on the $K$--band or $H$--band flux ratio as a proxy for mass ratio errors. 
We find that the mass ratio of the major merging pairs are in the range $1.2-2.3$ and therefore,
given the $1 \sigma$ uncertainties on stellar mass ratio estimates reported
in Table~\ref{tab_pairs}, none of the major merger pairs could be misclassified minor mergers. 
Taking into account the mass ratio between the two galaxies in a pair, and the 
separations $\Delta$v and $r_p$, we use the formalism of 
Kitzbichler and White (2008, formula 10) to compute the merging time--scale for each pair 
$T_{merg}$, as described in de Ravel et al. (2009). These timescales are compatible
with the Lotz et al. (2010) estimates, as discussed in Lopez--Sanjuan (2011). 
We list the main pair properties in Table \ref{tab_pairs}, including $T_{merg}$
and the redshift $z_{assembly}$ by which the pair would have merged into a 
single galaxy, taking into account the redshift of the pair and $T_{merg}$.
Images and spectra of each pair are presented in Figures \ref{pairs_1} to 
\ref{pairs_12}.
We describe the properties of each pair below.

{\bf COSMOS--511001467 A/B (Figure \ref{pairs_1}):} Redshifts of the two 
galaxies in this pair have been measured from two different VUDS observations
at $z_1=2.0970$ and $z_2=2.0961$, for a 
velocity difference of $\Delta v=87$ km s$^{-1}$.
The two galaxies are separated by 18.4 $h^{-1}$kpc, and are easily visible both 
in the HST/ACS F814W image and in the UltraVista near--infrared images. 
The most massive galaxy has a stellar mass of $0.3 \times 10^{10} M_{\sun}$ and 
the stellar mass ratio between the two galaxies is estimated to be $M_1/M_2=2$, 
hence a major merger. 
The South--West galaxy shows a compact nucleus surrounded by a faint diffuse 
component, while the North--Eastern galaxy is seen as two components in both 
HST/ACS F814W and UltraVista images. 
The spectra of the two galaxies have a UV flux rising to the blue up
to Ly$\alpha$, typical of star forming galaxies at these 
redshifts. 
Given the physical separation and mass difference, and following the 
prescription of Kitzbichler \& White (2008), this pair is expected to merge
within 2.3 Gyr.

{\bf COSMOS--510788270 A/B (Figure \ref{pairs_2}):} Two galaxies separated by 
15.4 $h^{-1}$kpc are at a redshift $z=2.9629$, with a velocity difference 
$\Delta v=38$ km s$^{-1}$, measured from two different VUDS observations. 
The most massive galaxy has a mass $0.7 \times 10^{10} M_{\sun}$ and the mass 
ratio between the two galaxies is estimated to be $M_1/M_2=1.5$, a major merger. 
One of the spectra shows Ly$\alpha$ in emission with 
$EW(Ly\alpha)_{rest}=15$\AA, while Ly$\alpha$ is observed in absorption for the 
other galaxy. Images are barely resolved at the HST/ACS resolution, showing 
a slightly extended and elliptical light distribution. 
This pair is expected to merge within 1.5 Gyr.

{\bf COSMOS--510175610/510778438 (Figure \ref{pairs_3}):} This pair at a 
redshift $z_1=3.0939$ is separated by 9.4$h^{-1}$kpc and $\Delta v=161$ 
km s$^{-1}$, measured in the same slit from VUDS observations. 
The brightest and most massive galaxy has a compact, although irregular, 
morphology, while the companion to the West, which is confirmed at the same 
redshift, is more diffuse and has a low surface brightness. 
Another companion is visible about 3$h^{-1}$kpc to the North--East,
and although its photometry is compatible with the redshift of the pair, no 
spectroscopic redshift information is available to confirm that it is physically 
linked to this system.
Both spectra have been obtained from the VUDS survey.
One of the VIMOS spectra shows a weak Ly$\alpha$ in emission, while the other 
is purely in absorption. The mass of the brightest galaxy is 
$5 \times 10^{10} M_{\sun}$, and the mass ratio is $M_1/M_2=6.3$,
considered to be a minor merger.
This pair is expected to merge within 0.7 Gyr.

{\bf VVDS--02h--520452183/520450423 (Figure \ref{pairs_4}):} The spectra of 
these two galaxies obtained from the same VVDS slit 
show a broad--line AGN with $z=1.8370$ and an absorption--line galaxy with 
$z=1.8333$, for a velocity difference $\Delta v=391$ km s$^{-1}$. 
This pair is separated by 22.3 $h^{-1}$kpc, and the estimated mass ratio is 
$M_1/M_2=1.9$, with the most massive galaxy, the AGN host, having a stellar 
mass $5.6 \times 10^{10} M_{\sun}$. 
This estimate is, however, quite uncertain given the presence of the AGN. 
Another object is observed in between the two main galaxies but no redshift 
information is available. The AGN host is compact, barely resolved at the 
seeing of the best CFHTLS image ($FWHM=0.6$ arcsec), while the companion galaxy
is slightly elongated and irregular in shape. 
Given the physical separation and mass difference, this pair is expected to 
merge within 1.1 Gyr, although this estimate is affected by the
uncertain mass estimate of the AGN host.

{\bf VVDS--02h--520478238/520478087 (Figure \ref{pairs_5}):} The two galaxies 
observed spectroscopically in the same VUDS slit have $z_1=2.2460$, 
$z_2=2.2463$ for a velocity difference $\Delta v=28$ km s$^{-1}$, and 
are separated by $r_p=25$ $h^{-1}$kpc. 
They are part of a group of four galaxies within
5 arcsec ($30 h^{-1}$kpc), with photometric redshifts compatible
with the redshift of the pair. 
The eastern component is the most massive with 
$M_{\star}=2.0 \times 10^{10} M_{\sun}$.
The mass ratio, $M_1/M_2=2$, indicates a major merger and results in an expected
time to merge within 1.8 Gyr.

{\bf VVDS--02h--910260902/910261083 (Figure \ref{pairs_6}):} The two galaxies, 
spectroscopically observed in the same VVDS slit with $z_1=2.3594$ and 
$z_2=2.3588$, are identified both in the CFHTLS $i$--band image and the 
$K$--band WIRDS image, with $r_p=12.3$ $h^{-1}$kpc and $\Delta v=54$ km s$^{-1}$ 
separation. The slit was placed on 
the centroid of the  blended CFHTLS image of the two galaxies, but still included
a significant fraction  of the flux of both galaxies to yield two well separated
spectra. 
The South--East galaxy is the most massive with $7.9 \times 10^{10} M_{\sun}$ 
and the mass ratio is $M_1/M_2=2.4$, the later indicating a major merger. 
While the massive galaxy shows a symmetric elongated shape, the North--West 
galaxy shows an irregular morphology, and the two are connected by a faint 
bridge.
Both galaxies show Ly$\alpha$ in emission, with $EW(Ly\alpha)_{rest}=50$\AA ~and 
$60$\AA, and integrated $Ly\alpha$ line flux $L_{Ly\alpha}=10^{42}$ 
and $3\times 10^{42}$ erg s$^{-1}$ ~respectively,
indicating strong star formation at the level of $1-2 M_{\sun}/yr$ (see e.g. Cassata et al. 2011). 
Given the observed separation of this pair, it is expected that it will merge 
within 0.6 Gyr.

{\bf VVDS--02h--910302317/910302929 (Figure \ref{pairs_7}):} The two main 
galaxies observed in spectroscopy in the same VVDS slit are respectively at 
$z_1=1.8171$, and $z_2=1.8154$, separated by 8.7 $h^{-1}$kpc along the 
East--West direction, and by $\Delta v=181$ km s$^{-1}$. 
The brightest / most massive galaxy has a mass 
$1.8 \times 10^{10} M_{\sun}$ and is made of two main components, well visible 
both in the $i$--band and $K$--band images.  
The mass ratio between the two galaxies in the pair is $M_1/M_2=1.7$, which
indicates a major merger.
This pair is expected to merge within 0.6 Gyr.

{\bf ECDFS--530034527 (Figure \ref{pairs_8}):} This is the highest redshift for 
which we have obtained a spectroscopic confirmation of a physical pair, with 
both 
galaxies at the same redshift $z=3.6500$, separated by 6.8 $h^{-1}$ kpc,
as measured from the same VUDS slit.
In Figure \ref{pairs_8} another companion is observed to the North--East, but 
there is no confirmation of its redshift. The stellar mass of the brightest and 
most massive galaxy is $1.2 \times 10^{10} M_{\sun}$, and the mass ratio with 
the companion is $M_1/M_2=1.3$, which indicates a major merger.  
The two spectra obtained in the VUDS survey show absorption line spectra, with 
only a weak $Ly\alpha$ emission for the brighter galaxy.
This pair is expected to merge within 0.5 Gyr.

{\bf ECDFS--530050663 (Figure \ref{pairs_9}):} This pair is made of 2 components
identified in the $H$--band CANDELS image, with redshifts $z_1=2.3250$ and 
$z_2=2.3244$ measured in the same VUDS slit. The stellar mass of the main component is 
$0.4 \times 10^{10} M_{\sun}$, and the mass ratio between components is 
$M_1/M_2=2$: a major merger. A third component is observed to the North--East, all three 
components being embedded in a low surface brightness emission. The pair 
identification is therefore ambiguous, as this configuration might be indeed a 
merger at an advanced stage surounded by tidal debris, or the result of three 
giant star forming regions in a single forming galaxy.
Under the merger hypothesis, the two main components are expected to merge
within 0.6 Gyr.  

{\bf ECDFS--530042814/530042840 (Figure \ref{pairs_10}):} The two galaxies are 
well--visible in the composite BVR MUSYC images and the $K$--band image.
These galaxies are at a mean redshift $z=2.9903$ with a velocity difference 
$\Delta v=278$ km s$^{-1}$ as measured in the same VUDS slit. 
These two main galaxies are separated by 6 $h^{-1}$kpc, and a third component 
is observed 6.5 $h^{-1}$kpc to the North--East. 
While the main galaxy shows Ly$\alpha$ in absorption, the other galaxy measured 
with VIMOS shows Ly$\alpha$ in emission, both spectra having a UV slope 
indicating strong star formation. The mass of the main galaxy is estimated to 
be $0.12 \times 10^{10} M_{\sun}$, and the mass ratio is $M_1/M_2=1.2$
almost an equal mass major merger. 
This pair is expected to merge withins 1.0 Gyr.

{\bf ECDFS--530046916 (Figure \ref{pairs_11}):} Two close but well separated 
images of  two galaxies are observed in both HST $z$--band (F850W) and 
$H$--band (F160W) images, with a redshift around $z=2.868$ measured in the 
same VUDS slit. 
The brightest / most massive galaxy has a regular morphology compatible with
a disc, while the companion galaxy has an irregular shape. 
The mass of the main galaxy is estimated to be 
$1.4 \times 10^{10} M_{\sun}$, and the mass ratio is $M_1/M_2=2.3$,
indicating a major merger. 
This pair will probably merge within 0.5 Gyr.

{\bf ECDFS--530036900/26931 (Figure \ref{pairs_12}):} Two well separated 
galaxies are observed in the HST $z$--band (F850W) and $H$--band (F160W) images 
from CANDELS, with 7.0 $h^{-1}$kpc and $\Delta v=78$ km s$^{-1}$ separation at a 
redshift of $z_1=3.3300$. One galaxy is quite compact, while the companion shows a 
sharp point--like component surrounded by a nebulous extension. 
The compact galaxy has been observed by VUDS and shows strong Ly$\alpha$
emission, while the fainter galaxy has been observed as part of the 
VIMOS--GOODS survey (Popesso et al. 2009). 
The mass of the main galaxy is estimated to be $1.3 \times 10^{10} M_{\sun}$, 
and the mass ratio is $M_1/M_2=1.2$, an almost equal mass major merger. 
This pair will probably merge in less than 0.6 Gyr.


   \begin{table*}
    \begin{center}
      \caption[]{Pair properties}
      \[
         \begin{array}{p{0.1\linewidth}lccccccccc}
           \hline \hline
            \noalign{\smallskip}
    Field     &  $Pair Galaxies$   & $Redshift$ & $Stellar Masses$^{(1)}    &  $Mass ratio$^{(2)}  & $Flux ratio$^{(3)}   & r_p^{(4)}      &  \Delta v^{(5)} & T_{merg}^{(6)} & z_{assembly}^{(7)} \\ 
              &  $ID numbers$      &   z_1      &     10^{10} M_{\sun} &                        &                       & $h$^{-1}$kpc$  &  $km s$^{-1}    & $Gyr$ & \\
            \noalign{\smallskip}
            \hline \hline
            \noalign{\smallskip}
    COSMOS    & 511001467 $a\&b$       &  2.0970  &   0.3 / 0.15  & 2.0 & 1.9\pm0.2  & 18.4\pm0.3    & 87   & 2.3 & 1.1 \\ 
              & 510788270 $a\&b$       &  2.9629  &   0.5 / 0.7   & 1.4 & 1.7\pm0.3  & 15.4\pm0.2    & 38   & 1.5 & 1.8 \\
              & 510175610 / 510778438  &  3.0939  &   0.8 / 5.0   & 6.3 & 4.2\pm0.2  &  9.4\pm0.2    & 161  & 0.7 & 2.3 \\
            \noalign{\smallskip}
            \hline
            \noalign{\smallskip}
    VVDS-02h  & 520452183 / 520450423  &  1.8370  &   5.6 / 3.0   & 1.9 &  2.2\pm0.1  & 22.3\pm1.2    & 390  & 1.1 & 1.3 \\ 
              & 520478238 / 520478087  &  2.2460  &   2.0 / 1.0   & 2.0 &  1.8\pm0.4  & 25.3\pm1.2    & 28   & 1.8 & 1.3 \\
              & 910260902 / 910261083  &  2.3594  &   3.4 / 7.9   & 2.3 &  2.0\pm0.1  & 12.3\pm1.1    & 54   & 0.6 & 1.9 \\
              & 910302317 / 910302929  &  1.8171  &   1.1 / 1.8   & 1.6 &  1.4\pm0.1  & 8.7\pm1.2     & 181  & 0.6 & 1.5 \\
            \noalign{\smallskip}
            \hline
            \noalign{\smallskip}
    ECDFS     & 530034527 $a\&b$       &  3.6500  &   1.2 / 0.9   & 1.3 & 1.5\pm0.1  & 6.6\pm0.2     & 0    & 0.5 & 2.9 \\ 
              & 530050663 $a\&b$       &  2.3250  &   0.4 / 0.2   & 2.0 & 1.9\pm0.2  & 5.5\pm0.3     & 54   & 0.6 & 1.9 \\
              & 530042814 / 530042840  &  2.9903  &   0.12 / 0.1  & 1.2 & 1.3\pm0.1  & 6.0\pm0.2     & 278  & 1.0 & 2.1 \\
              & 530046916 $a\&b$       &  2.8684  &   1.4 / 0.6   & 2.3 & 2.5\pm0.1  & 6.5\pm0.2     & 279  & 0.5 & 2.4 \\
              & 530036900 $a\&b$       &  3.3300  &   1.3 / 1.1   & 1.2 & 1.5\pm0.1  & 7.0\pm0.2     & 76   & 0.6 & 2.6 \\
            \noalign{\smallskip}
            \hline
         \end{array}
      \]
\caption{
$^{(1)}$ Stellar mass estimates from SED fitting; 
$^{(2)}$ Stellar mass ratio between the two galaxies in the pair, errors on the mass ratio have been 
computed from the error on the ratio of the H or K band flux measurements (see text);
$^{(3)}$ Flux ratio as measured on H or K-band
$^{(4)}$ $r_p$: transverse separation between the two galaxies in the pair,
errors in $r_p$ have been conservatively estimated
taking a one pixel error on the difference in the centroid measurement 
of each galaxy in the pair; 
$^{(5)}$ $\Delta v$: velocity separation along the line of sight; 
errors in $\Delta v$ are estimated to be $<100$ km s$^{-1}$ when the two redshifts are from the same slit
(10 pairs), and $\sim200$ km s$^{-1}$ when redshifts are from two different slits (2 pairs);
$^{(6)}$ $T_{merg}$: timescale for the pair to merge, using the Kitzbichler \& White (2008) prescription;
from errors in mass ratio, $r_p$ and $\Delta v$ errors on $T_{merg}$ are $\sim10-20$\%;
$^{(7)}$ $z_{assembly}$: redshift by which the two galaxies will have merged, obtained
combining the observed redshift and the $\Delta z$ corresponding to the merger timescale. 
}
\label{tab_pairs}
\end{center}
   \end{table*}


\section{Discussion and conclusions: major merging pairs at $2 \lesssim z<4$}
\label{discuss}

We have identified 12 pairs of galaxies with redshifts $1.81 \leq z \leq 3.65$ 
from the on--going VUDS and VVDS spectroscopic surveys. 
Both components of the pairs have a confirmed spectroscopic redshift obtained 
with VIMOS on the VLT and, therefore, comprise a unique sample of true physical 
pairs at these redshifts. 
The galaxies in our sample span a large mass range from $10^{9.1}M_{\sun}$ to 
$10^{11}M_{\sun}$.
The majority of galaxies in our pairs show signs of strong star formation, a common property at
these redshifts, as evidenced by either strong Ly$\alpha$ emission, strong UV continuum,
or both, and with one of them showing AGN activity.
The mass ratio of the merger is in the range $1 < M_1/M_2 < 6$, with 11 of 12 
pairs satisfying a major merger pair criterion $1 < M_1/M_2 < 4$.
The VUDS survey is still going--on, and deriving the survey selection
function is beyond the scope of this paper, however, we use these 
pairs to derive a preliminary estimate of the major merger pair fraction.
After correcting for the selection function and geometrical effects we derive a
first tentative measurement for the major merger fraction of  
$f_{merg} \sim (15-20)\%$ at $1.8 < z < 4.0$.
This value is comparable to the one observed from pair identification
at $1<z<1.8$ in the MASSIV survey (Lopez--Sanjuan et al. 2013), and 
measurements obtained from CAS morphological analysis up to $z=3$ 
(Conselice et al. 2003). 
Our merger fraction estimate will be further refined as the VUDS survey
proceeds.

Using the observed projected spatial and velocity separations, along with the 
prescription from Kitzbichler and White (2008), we have computed the merging 
timescale for each pair (Table \ref{tab_pairs}). 
With separations from 6 to $25 h^{-1}$kpc and mass ratio 1/6 to 1, the average 
merging timescale of our sample is $<T_{merg}>=1$ Gyr and the median is 0.7 Gyr.
Most of these pairs will therefore have merged before the peak of star formation 
at $z\simeq1.5$ (e.g. Cucciati et al. 2012). 
Given the average mass ratio of 1.75 for the 11 major merger pairs, the main 
galaxies involved in these mergers will have increased their 
stellar mass by $\sim60\%$ from $z\sim3$ to $z\sim1.5$ from the merging process 
alone. 

Our observations therefore provide unambiguous evidence of major merging
occurring at $2 \lesssim z<4$. It is clear that hierarchical assembly, 
with massive galaxies being built from the merging of less massive 
ones, is at work at these redshifts and that major merging is contributing to the 
assembly of mass in galaxies  at early times.
This mass assembly simply results from the sum of the masses in each galaxy in 
a merging pair, a simple and effective way to increase mass at each merging 
event. 
An additional increase of stellar mass from star formation triggered by the 
merging process is also possible, with a range of mass production 
identified in the literature, ranging from relatively large
(e.g. Kocevski et al. 2011) to more limited star bursts
(e.g. Mullaney et al. 2012), depending on the duration and strength of the
merger induced burst.
Merging is therefore a clear path to move low mass galaxies towards the higher 
end of the mass function, which contributes to the evolution of the stellar mass 
function (e.g. Ilbert et al. 2012; Ilbert et al. 2013). 
As identified by de Ravel et al. (2009) and Lopez--Sanjuan et al. (2013), 
galaxies in the local universe with a mass $10^{10.5}M_{\sun}$ will have 
assembled $30-40\%$ of their mass from major merging since $z\sim1.5$. 
Our results demonstrate that merging was also  contributing to mass assembly in 
galaxies at even earlier times.

These results are to be placed in the context
of the currently favored picture of galaxy assembly, with cold accretion
playing a key role in building--up mass in galaxies (e.g. Dekel et al. 2009). 
While cold accretion is observed in simulations as the primary mechanism fueling
galaxies with enough gas mass to sustain the strong, continuous star formation 
necessary to explain the star formation rate history, the  observational 
evidence for cold accretion at these redshifts remains elusive.  
While our results contribute to bring evidence that merging is at work,
it is important to note that merging is not exclusive to other mass assembly processes.
The high gas fraction of high redshift galaxies (Tacconi et al. 2010; 
Daddi et al. 2010) implies that other processes must also be at work, such as 
the continuous formation of stars potentially fueled by cold accretion.
The total mass growth of a galaxy must therefore be coming from all these
different processes at work. 
By precisely knowing the contribution of merging to global mass assembly, measured, 
e.g., by the growth of the stellar mass density,
one would ultimately be able to place upper limits to the total mass growth from
other processes, including cold gas accretion. 

The integrated contribution of merging processes to the complete history of 
mass assembly requires knowledge of the evolution of the merger rate since early
times (i.e., significantly beyond $z\sim2$). 
Building on the sample presented in this paper, the VUDS survey, when 
complete,  will enable a robust measurement of the merging rate out to $z\sim4$,
and an estimate of the total amount of mass assembled by the merging process 
since the early universe.


\begin{acknowledgements}
      This work is supported by
      funding from the European Research Council Advanced Grant 
      ERC--2010--AdG--268107--EARLY. 
\end{acknowledgements}



\end{document}